\documentclass[%
 reprint,
 superscriptaddress,
 amsmath,amssymb,
 aps,
]{revtex4-2}

\usepackage{graphicx}
\usepackage{dcolumn}
\usepackage{bm}
\usepackage[usenames,dvipsnames]{color} 
\usepackage{physics}
\usepackage{siunitx}
\usepackage{soul}
\usepackage{hyperref}
\usepackage{ulem} 
\usepackage{nicefrac}
\usepackage[all]{nowidow}

\begin{document}

\preprint{APS/123-QED}
\title{Hybridization of the amplitude mode in a confined fermionic superfluid}

\author{C.R. Cabrera}
\email{cesar.cabrera@physik.uni-hamburg.de}
\affiliation{Institute for Quantum Physics, Universität Hamburg, Luruper Chaussee 149, 22761 Hamburg, Germany}
\affiliation{The Hamburg Centre for Ultrafast Imaging, Universität Hamburg, Luruper Chaussee 149, 22761 Hamburg, Germany}

\author{R. Henke}
\affiliation{Institute for Quantum Physics, Universität Hamburg, Luruper Chaussee 149, 22761 Hamburg, Germany}

\author{L. Broers}
\affiliation{Center for Optical Quantum Technologies, Universität Hamburg, Luruper Chaussee 149, 22761 Hamburg, Germany}

\author{J. Skulte}
\author{H.P. Ojeda Collado}
\affiliation{The Hamburg Centre for Ultrafast Imaging, Universität Hamburg, Luruper Chaussee 149, 22761 Hamburg, Germany}
\affiliation{Center for Optical Quantum Technologies, Universität Hamburg, Luruper Chaussee 149, 22761 Hamburg, Germany}

\author{H. Biss}
\affiliation{Institute for Quantum Physics, Universität Hamburg, Luruper Chaussee 149, 22761 Hamburg, Germany}
\affiliation{The Hamburg Centre for Ultrafast Imaging, Universität Hamburg, Luruper Chaussee 149, 22761 Hamburg, Germany}

\author{L. Mathey}
\affiliation{Institute for Quantum Physics, Universität Hamburg, Luruper Chaussee 149, 22761 Hamburg, Germany}
\affiliation{The Hamburg Centre for Ultrafast Imaging, Universität Hamburg, Luruper Chaussee 149, 22761 Hamburg, Germany}
\affiliation{Center for Optical Quantum Technologies, Universität Hamburg, Luruper Chaussee 149, 22761 Hamburg, Germany}

\author{H. Moritz}
\email{hennning.moritz@physik.uni-hamburg.de}
\affiliation{Institute for Quantum Physics, Universität Hamburg, Luruper Chaussee 149, 22761 Hamburg, Germany}
\affiliation{The Hamburg Centre for Ultrafast Imaging, Universität Hamburg, Luruper Chaussee 149, 22761 Hamburg, Germany}

\date{\today}

\begin{abstract}

In phase transitions, spontaneous symmetry breaking results in a non-zero order parameter and two collective excitations: the Goldstone and the amplitude mode. 
These modes, which define key properties of superconductors and fermionic superfluids, are well-understood in homogeneous systems. 
However, their behavior under strong confinement remains largely unexplored, particularly when their excitation energy becomes comparable to the imposed discrete level spacing.
In this scenario, hybridization between different collective modes is expected to take place.
Here, we show how the amplitude mode hybridizes with a spatial mode in a confined fermionic superfluid.
Using lattice modulation spectroscopy, we observe the evolution of the mode throughout the entire crossover from the Bardeen-Cooper-Schrieffer (BCS) state to Bose-Einstein condensation (BEC) of molecules.
In the BCS regime, the excitation energy is located at twice the pairing gap, then gradually becomes an in-gap excitation in the strongly correlated regime.
Further to the BEC limit, the excitation energy approaches twice the level spacing. 
The spectral weight of this mode vanishes when approaching the superfluid critical temperature.
Our experimental results are in excellent agreement with an effective field theory, providing strong evidence that amplitude oscillations hybridize with and eventually transform into breathing oscillations of the order parameter.
The strong modification of the excitation spectrum reveals how confinement and finite-size effects impact fundamental modes of symmetry-broken states. 
\end{abstract}
\maketitle

Bardeen-Cooper-Schrieffer (BCS) theory \cite{Bardeen1957-ni} provides the framework for understanding superconductivity. 
It describes the formation of Cooper pairs and the emergence of a superconducting energy gap, $2\,\Delta$, corresponding to the minimum energy required to break them. 
However, as P.\ Anderson noted, significant deviations from this framework 
can occur in finite-size systems when the characteristic length or energy scale imposed by the confinement approaches the Cooper pair size or the pairing gap energy, respectively \cite{Anderson1959-wi}. 
These finite-size effects have been predicted and observed in systems such as nanowires and thin films \cite{Blatt1963-sq,PhysRevB.74.052502,PhysRevLett.17.632}, where confinement can modify the dimensionality, enhance the superfluid gap, or increase the critical temperature ($T_c$), see review~\cite{Bose2014-ay}. 
Such phenomena highlight how confinement can be used to engineer superconducting properties.

A key open question we address in this work is how confinement influences the fundamental collective modes of fermionic superfluids and superconductors. 
These systems are characterized by an order parameter, $\Psi = |\Psi| e^{-i\theta}$, supporting fluctuations in $\theta$ and $|\Psi|$, which correspond to the Goldstone mode (gapless phononic excitation) and the amplitude mode (gapped Higgs-like excitation at energy $2\Delta$), respectively \cite{Pekker2015-lk, Shimano2020-vg}.
While these collective modes are well understood in homogeneous systems, their behavior in the presence of confinement remains largely unexplored. 
For instance, the excitation spectrum is expected to change qualitatively \cite{Korolyuk2011-tu, Korolyuk2014-ix, Hertkorn2019-dn, Bjerlin2016-qt}, featuring a hybridization of oscillations of the amplitude and the spatial distribution of the order parameter.

\begin{figure}[tb!]
    \centering
    \includegraphics[width=\linewidth]{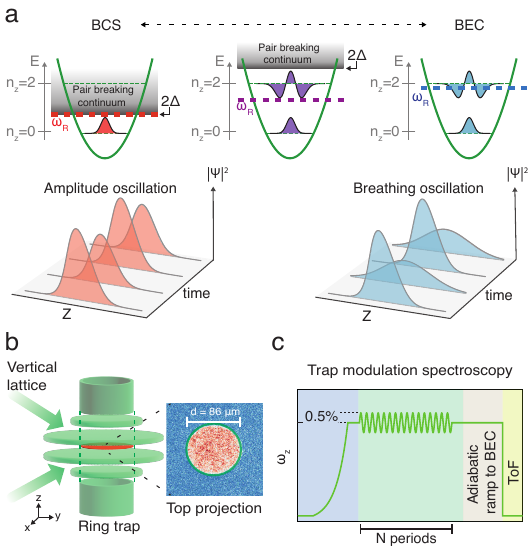}
    \caption{ 
    \textbf{a)} 
    Hybridization of amplitude oscillations with spatial excitations.
    In our confined fermionic superfluid, a well-defined excitation of the order parameter with frequency $\omega_\mathrm{R}$ is observed throughout the entire BEC-BCS crossover (dashed lines). 
    In the BCS regime, this excitation corresponds to a pure amplitude oscillation of the order parameter located at the onset of the pair-breaking continuum $2 \Delta$.
    This oscillation conserves the quantum number $n_z=0$, but the superfluid density $n_s = \int|\Psi|^2 dz$ oscillates in amplitude, illustrated with red curves.
    In the BEC regime, the excitation at $\omega_\mathrm{R}$ corresponds to a breathing mode with energy approaching $2 \hbar \omega_z$, involving the quantum numbers $n_z=0$ and $n_z=2$.
    This oscillation is sketched with blue curves.
    In the strongly interacting regime, when $\Delta$ becomes comparable to $\hbar \omega_z$, the resulting excitation is a hybridized mode with both breathing and amplitude character. 
    Here, the mode frequency drops below the pair-breaking continuum.
    \textbf{b)}
    We prepare a homogeneous Fermi gas in a ring-shaped box potential, strongly confined along $z$ with a repulsive lattice potential. The harmonic confinement is comparable to the Fermi energy $E_\mathrm{F}$ with $1.5 < E_\mathrm{F}/\hbar\omega_z <2$.
    \textbf{c)} 
    The trapping frequency $\omega_z$ is weakly modulated to excite spatial oscillations of the order parameter and amplitude oscillations via the corresponding modulation of the interaction strength.
    The deposited heat is measured by performing ToF after an adiabatic ramp to the BEC regime.
}
    \label{fig1}
\end{figure}

In previous work, evidence for amplitude oscillations of the order parameter has been reported in both superconductors \cite{Sooryakumar1980-bz,Matsunaga2014-ss,Measson14, Sherman2015-rm}, as well as in ultracold bosonic and fermionic superfluids \cite{Bissbort2011-nd,Endres2012-ci, Leonard2017-sq, Behrle2018-pn, Dyke2024-kx,Kell2024-jx}.
Additionally, a mesoscopic system has been used to explore precursors of the amplitude mode and its interplay with discrete energy shells \cite{Bayha2020-ek}. 
In contrast to these experiments, we study the effect of confinement on the amplitude mode in a macroscopic fermionic superfluid and observe strong evidence of its  hybridization with a collective mode of the trap.
This is achieved by exploiting the BEC-BCS crossover to tune the pairing gap energy, bringing it close to the trap level spacing.
The resulting hybridization is depicted in Fig.\,\ref{fig1}\,a.
In the BCS regime, the pairing gap energy is smaller than the trap level spacing.
Here, the system exhibits a well-defined resonance with frequency $\omega_R$ located at $2\Delta$ (Fig.\,\ref{fig1}\,a, left panel), where the superfluid density $n_s = \int|\Psi|^2 dz$ oscillates in amplitude (red curves).
Its frequency coincides with the onset of pair-breaking, however, it is a coherent excitation.
Towards the strongly interacting regime, the pairing gap increases, and the amplitude mode hybridizes with the nearest spatial mode of the trap with equal parity, specifically the quantum number $n_z=2$ (central panel). 
In this intermediate regime, the resonance energy $\omega_R$ drops below $2\Delta$, and the order parameter acquires a spatially oscillating component, admixing state $n_z=2$ to the ground state $n_z=0$.
Finally, deep in the BEC regime, the resonance approaches twice the trap level spacing (right panel), resembling a pure breathing mode (blue curves). 
Our interpretation is based on an effective field theory, which continuously connects the Gross-Pitaevski and Klein-Gordon functionals~\cite{Pekker2015-lk,Skulte2021}, showing remarkable agreement with our experimental results.

To engineer the mode hybridization, we prepare a homogeneous fermionic superfluid ($T^* < 0.02\, T_\mathrm{F}$) of $^6$Li atoms in the lowest two hyperfine states \cite{supmat,Hueck2018-cs}. 
The system is subjected to strong vertical confinement $\hbar \omega_z = h \cdot 8.7\,\mathrm{kHz}$ (Fig.\,\ref{fig1}\,b), comparable to the Fermi energy $E_\mathrm{F} = \hbar^2 4\pi n_{\mathrm{2D}}/2 m$~\cite{supmat}. 
Here, $T_F$ is the Fermi temperature,  $T^*$ is the temperature measured after an adiabatic sweep into the BEC regime \cite{supmat}, $m$ is the atom mass, $\hbar$ is the reduced Planck’s constant, and $n_{\mathrm{2D}}$ is the fermionic density per spin state. 
To reach the regime where the pairing energy is comparable to the vertical confinement energy, we tune the interaction parameter $\eta = \ln(k_{\mathrm{F}}a_{\mathrm{2D}})$ using a Feshbach resonance, where $k_\mathrm{F} = \sqrt{4 \pi n_{\mathrm{2D}}}$ is the Fermi momentum and $a_{\mathrm{2D}}$ is the 2D scattering length \cite{Petrov2001-kn,Levinsen2015-du,Turlapov2017-nl}.
To probe both spatial and amplitude oscillations of the order parameter, we vary the interaction strength \cite{Dyke2024-kx,Kell2024-jx} 
by employing trap modulation spectroscopy \cite{Bissbort2011-nd, Endres2012-ci}. 
This modulates the harmonic oscillator length $l_z$ and hence the scattering length $a_{\mathrm{2D}}$~\cite{supmat}.

Specifically, we modulate the trapping frequency with an amplitude $\alpha \leq 0.5\,\%$ at a frequency $\omega_{\textrm{m}}$ for a fixed number of periods, $N = 240$,  followed by a rethermalization time of $20\,\mathrm{ms}$.
To obtain the deposited energy, we adiabatically ramp into the BEC regime ($\eta = -1.7 $) and measure the bimodal momentum distribution after time-of-flight (ToF). 
Here, we extract the condensate amplitude $A(\omega_{\textrm{m}})$. 
This is compared to the condensate amplitude of the unperturbed system $A(0)$, defining our response function $R(\omega_{\textrm{m}})=A(0)/A(\omega_\textrm{m})-1$ \cite{supmat}.
This normalized function deviates from zero when energy is deposited in the system, revealing resonant frequencies.  

\begin{figure}[t!]
    \centering
    \includegraphics[width=\linewidth]{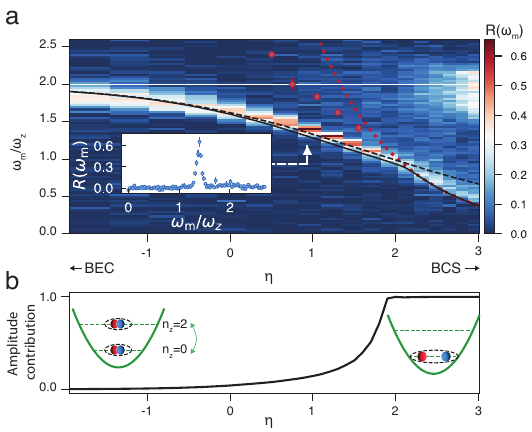}
    \caption{
    \textbf{a)} 
    Spectral response $R(\omega_\mathrm{m})$ 
    of a confined fermionic superfluid throughout the BEC-BCS crossover, measured via trap modulation spectroscopy.
    The inset shows a typical response $R(\omega_\mathrm{m})$ for $\eta = 0.97$, with
    a resonance located below the single-particle excitation energy $2\hbar \omega_z$. 
    In the BCS limit ($\eta \gg 0$), the energy of the resonance agrees with the mean-field value $2\Delta$ (red dotted line).
    At $\eta =2$, it deviates from $2 \Delta$, evolving into an in-gap excitation. 
    This is supported by an experimental measurement of $2\Delta$ using Bragg spectroscopy (red points). 
    In the BEC regime ($\eta \ll 0 $), where no amplitude mode exists, the excitation corresponds to a breathing mode along the confined direction. 
    Here, its energy is in good agreement with an analytical calculation of the energy difference between the $n_z=0$ and $n_z=2$ eigenmodes (black dashed line) extracted from an effective field theory (see main text). 
    The full numerical simulation shows excellent agreement in the entire crossover (black solid line). 
    The error bars show the standard deviation of the mean.
    \textbf{b)} 
    Hybridization of the amplitude mode. For values $\eta > 1.9$, the amplitude character is dominant, quantified by the amplitude contribution extracted from the effective field theory (black line). Here, the excitation does not conserve the total superfluid density, while conserving the quantum number $n_z=0$.
    In the BEC regime, the superfluid density is conserved and oscillates spatially involving the $n_z = 2$ eigenmode. 
    In the crossover, the composition of the collective mode corresponds to both amplitude and spatial oscillations of the order parameter.
    }
    \label{fig2}
\end{figure}

In the first series of experiments, we start with a fermionic density of $n_{\mathrm{2D}} \approx \SI{1.5}{atoms/\micro m^2}$, resulting in a ratio of $E_\mathrm{F}/\hbar \omega_z = 1.8$.
A typical response from trap modulation is shown in the inset of Fig.\,\ref{fig2}\,a.
This measurement is extended throughout the entire BEC-BCS crossover in Fig.\,\ref{fig2}\,a, where the color encodes the response $R(\omega_\mathrm{m})$. 
The vertical axis is normalized to the trapping frequency $\omega_z$, which is independently determined  using a fermionic gas with negligible interactions. 
The system exhibits two distinct excitations.
Within the BCS regime, one appears as a broad resonance located at $\omega_{\textrm{m}} \approx 2\omega_{z}$, corresponding to single-particle excitations of unpaired atoms to the energy level $n_z=2$. 
The other manifests as a narrower resonance at lower frequency, gradually increasing from the BCS ($\eta =3$) to the BEC $(\eta =-2$) regime, eventually approaching $2\omega_z$. 

In the weakly interacting BCS regime, the well-defined resonance is located at nearly twice the pairing gap $\Delta$. 
It agrees with the mean-field prediction of $\Delta = \sqrt{2 E_\mathrm{b} E_\mathrm{F}}$ (red dotted line), where the binding energy $E_\mathrm{b}$ is calculated in the 2D limit \cite{supmat}. 
This agreement breaks down towards the strongly interacting regime for $\eta < 1.9$, where the observed mode deviates from the onset of the pair-breaking continuum, measured independently via Bragg spectroscopy (red points in Fig.\,\ref{fig2}\,a) \cite{supmat,Veeravalli2008-hl,Hoinka2017-qp,Sobirey2021-xy,Biss2022-ra}, and instead becomes an in-gap excitation. 
Finally, in the BEC regime, the order parameter is directly related to the total density and hence only supports spatial oscillations along the vertical confinement \cite{Pekker2015-lk}. 
Here, the resonance corresponds to a breathing excitation of the condensate shifted downwards from $2\hbar \omega_z$ due to repulsive dimer-dimer interactions.

To understand the continuous evolution and character of the resonance, we develop a theoretical model in which the dynamics of the order parameter $\Psi(z,t)$ are described by the equation of motion
\begin{multline}
i K_1 \frac{\partial \Psi(z,t)}{\partial t} - K_2 \frac{\partial^2 \Psi(z,t)}{\partial t^2} \\ 
= \left(-\frac{\hbar^2\partial^2_z}{2 M} + V(z,t) - r_0 + u_0|\Psi(z,t)|^2\right) \Psi(z,t),
\label{eq:1}
\end{multline}
derived from a zero-temperature effective field theory \cite{supmat}.
We consider a homogeneous system along the x-y plane and decompose $\Psi(z,t)$ into the $n_z=0$ and $n_z=2$ eigenstates of the harmonic potential $V(z,t)$ \cite{supmat}.
In this model, the coefficients $K_1$, $K_2$, $r_0$, and $u_0$ parameterize the BEC-BCS crossover.
These are determined from theoretical predictions for the energy gap, condensate density \cite{Salasnich2007}, and speed of sound \cite{Shi2015} in the crossover \cite{supmat}. 
In the limit where $K_1 = 0$, Eq.\,\ref{eq:1} reduces to a nonlinear Klein-Gordon equation, describing the order parameter dynamics in the weakly interacting BCS regime.
The resulting equation is Lorentz invariant which is a manifestation of particle-hole symmetry. 
In the BEC limit, where  $K_2 = 0$, Eq.\,\ref{eq:1} reduces to the Gross-Pitaevskii equation. 

The black dashed line in Fig.\,\ref{fig2}\,a represents the eigenfrequency difference between the ground and second excited state as analytically obtained from Eq.\,\ref{eq:1}, taking into account interactions to leading order \cite{supmat}. 
Towards the BEC regime, this analytical result is in excellent agreement with the experimental observation.
The full numerical simulation in the presence of trap modulation (black solid line) extends the agreement throughout the entire BEC-BCS crossover without any fitting parameters.
Beyond capturing the mode frequencies, this numerical approach also reveals the composition of the collective excitation due to hybridization, distinguishing between its amplitude and breathing mode components. 
The theoretical amplitude mode contribution is shown in Fig.\,\ref{fig2}\,b, and represents the fraction of the response where the total superfluid density oscillates in time ~\cite{supmat}.
In the BCS regime ($\eta \gtrsim 1.9$), this contribution approaches its maximum value, resembling a pure amplitude mode (see red curves in Fig.\,\ref{fig1}\,a).
Towards the strongly correlated regime ($\eta \sim 1$), the amplitude mode contribution gradually decreases, and the mode gains breathing character.
Finally, in the BEC limit, the dominant excitation is a pure breathing mode, where the superfluid density is constant in time, and oscillates spatially between $n_z=0$ and $n_z=2$ (see blue curves in Fig.\,\ref{fig1}\,a).

\begin{figure}[t!]
    \centering
    \includegraphics[width=\linewidth]{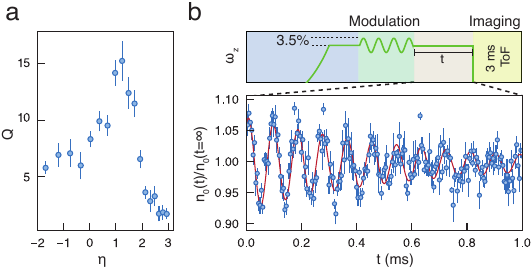}
    \caption{
    \textbf{a)} 
    Quality factor along the BEC-BCS crossover. 
    We quantify the spectral width via the quality factor $Q$, extracted from the spectral response in the BEC-BEC crossover.
    For all interaction strengths we find $Q>1$, signalling a long-lived excitation throughout the BEC-BCS crossover.
    The maximum value $Q=15 \pm 2$ occurs in the strongly correlated regime.
    \textbf{b)}
    In this regime ($\eta = 1.2$), the collective mode is resonantly excited with a four-period modulation of the trapping frequency. 
    This is followed by a free evolution for a time span $t$. 
    We observe oscillations of the central density $n_0$ after a ToF expansion of \SI{3}{\ms}.
    The experimental data is fitted with a damped oscillation yielding a decay time $\tau = (610 \pm 60)\, \mu\mathrm{s}$ which results in $Q = 21 \pm 2$.
    } 
    \label{fig3}
\end{figure}  

One feature not captured by our theoretical model is the behavior of the spectral width throughout the crossover, particularly its narrowing in the hybridization regime, which suggests a long-lived and well-defined mode. 
We quantify the coherence of the mode by computing the quality factor $Q = \omega_0/\delta \omega$, where $\omega_0$ is the resonance frequency and $\delta \omega$ the full width at half maximum extracted from the spectrum.
In the entire crossover, the mode remains underdamped with $Q > 0.5$ (Fig.\,\ref{fig3}\,a).
Notably, the largest quality factors are found at $\eta < 1.9$, reaching a maximum value in the strongly interacting regime ($\eta \sim 1$).
Here, the mode energy is below the onset of the pair-breaking continuum, suggesting that single-particle decay is suppressed, leading to longer lifetimes \cite{Korolyuk2014-ix,Verresen2019-jp,Lorenzana24}. 
This contrasts with a conventional case, where the amplitude mode resides at the onset of the pair-breaking continuum, resulting in strong damping and limited visibility \cite{Benfatto15,Benfatto16,Kurkjian2019-ng,Kurkjian2020-xt}.
To further confirm the long-lived nature of the mode, we track the free coherent oscillations of the system at $\eta = 1.2$ after a short resonant excitation (Fig.\,\ref{fig3}\,b). 
We observe a long-lived oscillation at the collective mode frequency $\omega_0$ when measuring the change in the central density of the cloud after \SI{3}{\ms} of ToF \cite{supmat}.
Fitting the data with an exponentially damped oscillation with a decay time $\tau$ \cite{supmat}, we obtain a quality factor $Q = \tau \omega_0 / 2 \approx 21 \pm 2$, comparable with our results from trap modulation spectroscopy. 

\begin{figure}[bt!]
    \centering
    \includegraphics[width=\linewidth]{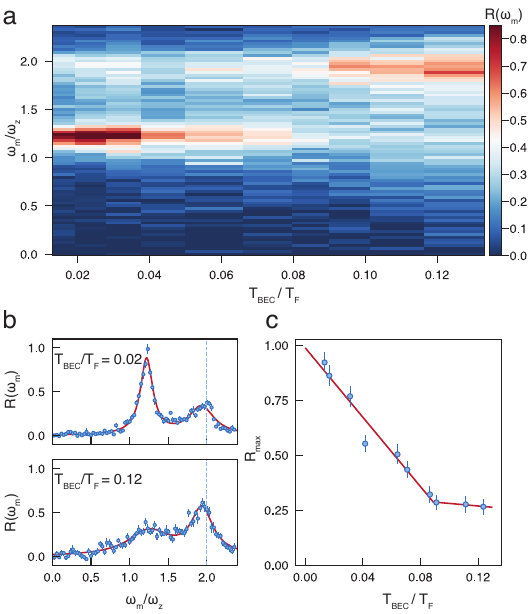}
    \caption{
    \textbf{a)}
    Vanishing of the collective mode above the critical temperature.
    We measure the spectral response $R(\omega_\mathrm{m})$ in the strongly correlated regime ($\eta=1.2$) as a function of temperature. 
    The collective mode frequency remains nearly constant at $\omega_\mathrm{m}/\omega_z \sim 1.2$.
    As the temperature increases, the mode broadens and its visibility diminishes, signaling the loss of coherence.
    Above a critical temperature, single-particle transitions at 2$\hbar \omega_z$ become the dominant process due to thermally broken pairs.
    \textbf{b)}
    To extract the critical temperature, we fit the maximum response of the collective mode $R_{\mathrm{max}}$ by a double Lorentzian function (red line) shown for two vertical cuts of panel a.  
    \textbf{c)}
    The temperature dependence of the maximum response shows a kink, which we identify as the critical temperature $T^*_{\mathrm{c}}$. 
    Employing a bilinear fit (red line), we obtain  $T^*_{\mathrm{c}} = 0.088(4)\,T_\mathrm{F}$.
    }
    \label{fig4}
\end{figure}

Since the observed dynamics are compatible with oscillations of the order parameter, we expect a strong temperature dependence, especially close to the superfluid critical temperature $T_c$. 
To explore this further, we prepare a strongly interacting system ($\eta = 1.2$) at different temperatures and perform trap modulation spectroscopy \cite{supmat}.
Fig.\,\ref{fig4}\,a shows the spectral response as a function of temperature $T^*/T_\mathrm{F}$.
With increasing temperature, the resonance at $2\hbar\omega_z$ associated with single-particle excitations from thermally broken pairs becomes the dominant contribution to the spectrum (see Fig.\,\ref{fig4}\,b).
Meanwhile, the collective mode with lower frequency broadens and its spectral weight decreases.
The frequency remains constant, resembling the behavior previously observed in Refs.\ ~\cite{Vaswani2021-mw, Dyke2024-kx,Kell2024-jx}. 
In Fig.\,\ref{fig4}\,c, the  peak amplitude of the mode as a function of temperature is plotted.
The observed trend is well described by a bilinear fit, featuring a kink at a temperature of $T^*_{\mathrm{c}}= 0.088(4)\,T_\mathrm{F}$.
Remarkably, our previous experiments on the critical velocity in 2D Fermi gases show a critical temperature of $T^*_{\mathrm{c}} = 0.09(2)\,T_\mathrm{F}$~\cite{Sobirey2021-xy}.
This supports the interpretation that the collective mode is an excitation of the order parameter which vanishes above a critical temperature.

In conclusion, our study focuses on a fundamental feature of strongly confined fermionic superfluids: the hybridization of collective modes. 
We observe a well-defined resonance in the entire BEC-BCS crossover, which we interpret as an excitation of the order parameter. 
This interpretation is supported by the strong suppression of the resonance above the critical temperature and the excellent agreement with the effective field theory. 

Additionally, we show that trap modulation spectroscopy is a powerful tool for determining the pairing gap in the BCS regime and the superfluid-to-normal transition in the crossover. This is of interest for studying spin-imbalanced superfluids and other strongly correlated 2D systems.

Beyond ultracold atoms, these results provide insights into the hybridization of modes in confined fermionic superfluids and superconductors, such as low-dimensional materials and nano-superconducting devices. 
Here, the hybridization could influence fundamental properties such as the critical velocity or critical temperature. 
Our work paves the way to investigate more complex dynamics of the order parameter under confinement~\cite{Hannibal2015,Hannibal2018}, such as frequency beating~\cite{HP2023}, drum modes \cite{Endres2012-ci}, exotic Higgs bound states \cite{Nakayama2015-ro,Nakayama2019-xy} or Leggett modes \cite{Legget66,Krull16,Lara19}. 
Finally, it is an exciting prospect to use confinement to control the properties of many-body systems by admixing higher harmonic oscillator states \cite{PhysRevX.13.021013,PhysRevA.110.L051302}.

\textbf{Acknowledgements.}
We gratefully acknowledge fruitful discussions with Y.\ Castin, T.\ Enss, D.\ Jaksch, J.\ Leonard, J.\ Levinsen, T.\ Lompe, M.\ Parish, P. Pieri and M.\ Zwierlein.
This work is supported by the Deutsche Forschungsgemeinschaft (DFG, German Research Foundation) in the framework of SFB-925—project 170620586—and the Cluster of Excellence “CUI: Advanced Imaging of Matter”—EXC 2056—project ID 390715994 and co-financed by ERDF of the European Union and by “Fonds of the Hamburg Ministry of Science, Research, Equalities and Districts (BWFGB)”.

\textbf{Author Contributions.} 
R.\,H., H.\,B.\ and C.\,R.\,C.\ performed the experiments and analysed the data.
L.\,B., J.\,S.\ and H.\,P.\,O.\,C.\ developed the theoretical model and numerical simulations.
L.\,M.\ and H.\,M.\ supervised the work. 
All authors contributed extensively to the interpretation of the results and to the preparation of the manuscript.

\textbf{Data availability.}
The datasets supporting this study are available from the corresponding authors upon request.

\textbf{Competing interests.} The authors declare no competing interests.

\textbf{Correspondence and requests for materials} should be addressed to C. R. Cabrera or
H. Moritz.

\newpage
\bibliography{Bibliography}

\begin{thebibliography}{59}%
\makeatletter
\providecommand \@ifxundefined [1]{%
 \@ifx{#1\undefined}
}%
\providecommand \@ifnum [1]{%
 \ifnum #1\expandafter \@firstoftwo
 \else \expandafter \@secondoftwo
 \fi
}%
\providecommand \@ifx [1]{%
 \ifx #1\expandafter \@firstoftwo
 \else \expandafter \@secondoftwo
 \fi
}%
\providecommand \natexlab [1]{#1}%
\providecommand \enquote  [1]{``#1''}%
\providecommand \bibnamefont  [1]{#1}%
\providecommand \bibfnamefont [1]{#1}%
\providecommand \citenamefont [1]{#1}%
\providecommand \href@noop [0]{\@secondoftwo}%
\providecommand \href [0]{\begingroup \@sanitize@url \@href}%
\providecommand \@href[1]{\@@startlink{#1}\@@href}%
\providecommand \@@href[1]{\endgroup#1\@@endlink}%
\providecommand \@sanitize@url [0]{\catcode `\\12\catcode `\$12\catcode
  `\&12\catcode `\#12\catcode `\^12\catcode `\_12\catcode `\%12\relax}%
\providecommand \@@startlink[1]{}%
\providecommand \@@endlink[0]{}%
\providecommand \url  [0]{\begingroup\@sanitize@url \@url }%
\providecommand \@url [1]{\endgroup\@href {#1}{\urlprefix }}%
\providecommand \urlprefix  [0]{URL }%
\providecommand \Eprint [0]{\href }%
\providecommand \doibase [0]{https://doi.org/}%
\providecommand \selectlanguage [0]{\@gobble}%
\providecommand \bibinfo  [0]{\@secondoftwo}%
\providecommand \bibfield  [0]{\@secondoftwo}%
\providecommand \translation [1]{[#1]}%
\providecommand \BibitemOpen [0]{}%
\providecommand \bibitemStop [0]{}%
\providecommand \bibitemNoStop [0]{.\EOS\space}%
\providecommand \EOS [0]{\spacefactor3000\relax}%
\providecommand \BibitemShut  [1]{\csname bibitem#1\endcsname}%
\let\auto@bib@innerbib\@empty
\bibitem [{\citenamefont {Bardeen}\ \emph {et~al.}(1957)\citenamefont
  {Bardeen}, \citenamefont {Cooper},\ and\ \citenamefont
  {Schrieffer}}]{Bardeen1957-ni}%
  \BibitemOpen
  \bibfield  {author} {\bibinfo {author} {\bibfnamefont {J.}~\bibnamefont
  {Bardeen}}, \bibinfo {author} {\bibfnamefont {L.~N.}\ \bibnamefont
  {Cooper}},\ and\ \bibinfo {author} {\bibfnamefont {J.~R.}\ \bibnamefont
  {Schrieffer}},\ }\bibfield  {title} {\bibinfo {title} {Theory of
  superconductivity},\ }\href {https://doi.org/10.1103/PhysRev.108.1175}
  {\bibfield  {journal} {\bibinfo  {journal} {Phys. Rev.}\ }\textbf {\bibinfo
  {volume} {108}},\ \bibinfo {pages} {1175} (\bibinfo {year}
  {1957})}\BibitemShut {NoStop}%
\bibitem [{\citenamefont {Anderson}(1959)}]{Anderson1959-wi}%
  \BibitemOpen
  \bibfield  {author} {\bibinfo {author} {\bibfnamefont {P.~W.}\ \bibnamefont
  {Anderson}},\ }\bibfield  {title} {\bibinfo {title} {Theory of dirty
  superconductors},\ }\href@noop {} {\bibfield  {journal} {\bibinfo  {journal}
  {J. Phys. Chem. Solids}\ }\textbf {\bibinfo {volume} {11}},\ \bibinfo {pages}
  {26} (\bibinfo {year} {1959})}\BibitemShut {NoStop}%
\bibitem [{\citenamefont {Blatt}\ and\ \citenamefont
  {Thompson}(1963)}]{Blatt1963-sq}%
  \BibitemOpen
  \bibfield  {author} {\bibinfo {author} {\bibfnamefont {J.~M.}\ \bibnamefont
  {Blatt}}\ and\ \bibinfo {author} {\bibfnamefont {C.~J.}\ \bibnamefont
  {Thompson}},\ }\bibfield  {title} {\bibinfo {title} {Shape resonances in
  superconducting thin films},\ }\href@noop {} {\bibfield  {journal} {\bibinfo
  {journal} {Phys. Rev. Lett.}\ }\textbf {\bibinfo {volume} {10}},\ \bibinfo
  {pages} {332} (\bibinfo {year} {1963})}\BibitemShut {NoStop}%
\bibitem [{\citenamefont {Shanenko}\ \emph {et~al.}(2006)\citenamefont
  {Shanenko}, \citenamefont {Croitoru}, \citenamefont {Zgirski}, \citenamefont
  {Peeters},\ and\ \citenamefont {Arutyunov}}]{PhysRevB.74.052502}%
  \BibitemOpen
  \bibfield  {author} {\bibinfo {author} {\bibfnamefont {A.~A.}\ \bibnamefont
  {Shanenko}}, \bibinfo {author} {\bibfnamefont {M.~D.}\ \bibnamefont
  {Croitoru}}, \bibinfo {author} {\bibfnamefont {M.}~\bibnamefont {Zgirski}},
  \bibinfo {author} {\bibfnamefont {F.~M.}\ \bibnamefont {Peeters}},\ and\
  \bibinfo {author} {\bibfnamefont {K.}~\bibnamefont {Arutyunov}},\ }\bibfield
  {title} {\bibinfo {title} {Size-dependent enhancement of superconductivity in
  al and sn nanowires: Shape-resonance effect},\ }\href
  {https://doi.org/10.1103/PhysRevB.74.052502} {\bibfield  {journal} {\bibinfo
  {journal} {Phys. Rev. B}\ }\textbf {\bibinfo {volume} {74}},\ \bibinfo
  {pages} {052502} (\bibinfo {year} {2006})}\BibitemShut {NoStop}%
\bibitem [{\citenamefont {Abeles}\ \emph {et~al.}(1966)\citenamefont {Abeles},
  \citenamefont {Cohen},\ and\ \citenamefont {Cullen}}]{PhysRevLett.17.632}%
  \BibitemOpen
  \bibfield  {author} {\bibinfo {author} {\bibfnamefont {B.}~\bibnamefont
  {Abeles}}, \bibinfo {author} {\bibfnamefont {R.~W.}\ \bibnamefont {Cohen}},\
  and\ \bibinfo {author} {\bibfnamefont {G.~W.}\ \bibnamefont {Cullen}},\
  }\bibfield  {title} {\bibinfo {title} {Enhancement of superconductivity in
  metal films},\ }\href {https://doi.org/10.1103/PhysRevLett.17.632} {\bibfield
   {journal} {\bibinfo  {journal} {Phys. Rev. Lett.}\ }\textbf {\bibinfo
  {volume} {17}},\ \bibinfo {pages} {632} (\bibinfo {year} {1966})}\BibitemShut
  {NoStop}%
\bibitem [{\citenamefont {Bose}\ and\ \citenamefont
  {Ayyub}(2014)}]{Bose2014-ay}%
  \BibitemOpen
  \bibfield  {author} {\bibinfo {author} {\bibfnamefont {S.}~\bibnamefont
  {Bose}}\ and\ \bibinfo {author} {\bibfnamefont {P.}~\bibnamefont {Ayyub}},\
  }\bibfield  {title} {\bibinfo {title} {A review of finite size effects in
  quasi-zero dimensional superconductors},\ }\href@noop {} {\bibfield
  {journal} {\bibinfo  {journal} {Rep. Prog. Phys.}\ }\textbf {\bibinfo
  {volume} {77}},\ \bibinfo {pages} {116503} (\bibinfo {year}
  {2014})}\BibitemShut {NoStop}%
\bibitem [{\citenamefont {Pekker}\ and\ \citenamefont
  {Varma}(2015)}]{Pekker2015-lk}%
  \BibitemOpen
  \bibfield  {author} {\bibinfo {author} {\bibfnamefont {D.}~\bibnamefont
  {Pekker}}\ and\ \bibinfo {author} {\bibfnamefont {C.~M.}\ \bibnamefont
  {Varma}},\ }\bibfield  {title} {\bibinfo {title} {{Amplitude/Higgs} modes in
  condensed matter physics},\ }\href
  {https://doi.org/10.1146/annurev-conmatphys-031214-014350} {\bibfield
  {journal} {\bibinfo  {journal} {Annu. Rev. Condens. Matter Phys.}\ }\textbf
  {\bibinfo {volume} {6}},\ \bibinfo {pages} {269} (\bibinfo {year}
  {2015})}\BibitemShut {NoStop}%
\bibitem [{\citenamefont {Shimano}\ and\ \citenamefont
  {Tsuji}(2020)}]{Shimano2020-vg}%
  \BibitemOpen
  \bibfield  {author} {\bibinfo {author} {\bibfnamefont {R.}~\bibnamefont
  {Shimano}}\ and\ \bibinfo {author} {\bibfnamefont {N.}~\bibnamefont
  {Tsuji}},\ }\bibfield  {title} {\bibinfo {title} {{H}iggs mode in
  superconductors},\ }\href
  {https://doi.org/10.1146/annurev-conmatphys-031214-014350} {\bibfield
  {journal} {\bibinfo  {journal} {Annu. Rev. Condens. Matter Phys.}\ }\textbf
  {\bibinfo {volume} {11}},\ \bibinfo {pages} {103} (\bibinfo {year}
  {2020})}\BibitemShut {NoStop}%
\bibitem [{\citenamefont {Korolyuk}\ \emph {et~al.}(2011)\citenamefont
  {Korolyuk}, \citenamefont {Kinnunen},\ and\ \citenamefont
  {T{\"o}rm{\"a}}}]{Korolyuk2011-tu}%
  \BibitemOpen
  \bibfield  {author} {\bibinfo {author} {\bibfnamefont {A.}~\bibnamefont
  {Korolyuk}}, \bibinfo {author} {\bibfnamefont {J.~J.}\ \bibnamefont
  {Kinnunen}},\ and\ \bibinfo {author} {\bibfnamefont {P.}~\bibnamefont
  {T{\"o}rm{\"a}}},\ }\bibfield  {title} {\bibinfo {title} {Density response of
  a trapped {F}ermi gas: A crossover from the pair vibration mode to the
  {G}oldstone mode},\ }\href {https://doi.org/10.1103/PhysRevA.84.033623}
  {\bibfield  {journal} {\bibinfo  {journal} {Phys. Rev. A}\ }\textbf {\bibinfo
  {volume} {84}},\ \bibinfo {pages} {033623} (\bibinfo {year}
  {2011})}\BibitemShut {NoStop}%
\bibitem [{\citenamefont {Korolyuk}\ \emph {et~al.}(2014)\citenamefont
  {Korolyuk}, \citenamefont {Kinnunen},\ and\ \citenamefont
  {T{\"o}rm{\"a}}}]{Korolyuk2014-ix}%
  \BibitemOpen
  \bibfield  {author} {\bibinfo {author} {\bibfnamefont {A.}~\bibnamefont
  {Korolyuk}}, \bibinfo {author} {\bibfnamefont {J.~J.}\ \bibnamefont
  {Kinnunen}},\ and\ \bibinfo {author} {\bibfnamefont {P.}~\bibnamefont
  {T{\"o}rm{\"a}}},\ }\bibfield  {title} {\bibinfo {title} {Collective
  excitations of a trapped {F}ermi gas at finite temperature},\ }\href
  {https://doi.org/10.1103/PhysRevA.89.013602} {\bibfield  {journal} {\bibinfo
  {journal} {Phys. Rev. A}\ }\textbf {\bibinfo {volume} {89}},\ \bibinfo
  {pages} {013602} (\bibinfo {year} {2014})}\BibitemShut {NoStop}%
\bibitem [{\citenamefont {Hertkorn}\ \emph {et~al.}(2019)\citenamefont
  {Hertkorn}, \citenamefont {B{\"o}ttcher}, \citenamefont {Guo}, \citenamefont
  {Schmidt}, \citenamefont {Langen}, \citenamefont {B{\"u}chler},\ and\
  \citenamefont {Pfau}}]{Hertkorn2019-dn}%
  \BibitemOpen
  \bibfield  {author} {\bibinfo {author} {\bibfnamefont {J.}~\bibnamefont
  {Hertkorn}}, \bibinfo {author} {\bibfnamefont {F.}~\bibnamefont
  {B{\"o}ttcher}}, \bibinfo {author} {\bibfnamefont {M.}~\bibnamefont {Guo}},
  \bibinfo {author} {\bibfnamefont {J.~N.}\ \bibnamefont {Schmidt}}, \bibinfo
  {author} {\bibfnamefont {T.}~\bibnamefont {Langen}}, \bibinfo {author}
  {\bibfnamefont {H.~P.}\ \bibnamefont {B{\"u}chler}},\ and\ \bibinfo {author}
  {\bibfnamefont {T.}~\bibnamefont {Pfau}},\ }\bibfield  {title} {\bibinfo
  {title} {Fate of the amplitude mode in a trapped dipolar supersolid},\ }\href
  {https://doi.org/10.1103/PhysRevLett.123.193002} {\bibfield  {journal}
  {\bibinfo  {journal} {Phys. Rev. Lett.}\ }\textbf {\bibinfo {volume} {123}},\
  \bibinfo {pages} {193002} (\bibinfo {year} {2019})}\BibitemShut {NoStop}%
\bibitem [{\citenamefont {Bjerlin}\ \emph {et~al.}(2016)\citenamefont
  {Bjerlin}, \citenamefont {Reimann},\ and\ \citenamefont
  {Bruun}}]{Bjerlin2016-qt}%
  \BibitemOpen
  \bibfield  {author} {\bibinfo {author} {\bibfnamefont {J.}~\bibnamefont
  {Bjerlin}}, \bibinfo {author} {\bibfnamefont {S.~M.}\ \bibnamefont
  {Reimann}},\ and\ \bibinfo {author} {\bibfnamefont {G.~M.}\ \bibnamefont
  {Bruun}},\ }\bibfield  {title} {\bibinfo {title} {{Few-Body} precursor of the
  higgs mode in a fermi gas},\ }\href
  {https://doi.org/10.1103/PhysRevLett.116.155302} {\bibfield  {journal}
  {\bibinfo  {journal} {Phys. Rev. Lett.}\ }\textbf {\bibinfo {volume} {116}},\
  \bibinfo {pages} {155302} (\bibinfo {year} {2016})}\BibitemShut {NoStop}%
\bibitem [{\citenamefont {Sooryakumar}\ and\ \citenamefont
  {Klein}(1980)}]{Sooryakumar1980-bz}%
  \BibitemOpen
  \bibfield  {author} {\bibinfo {author} {\bibfnamefont {R.}~\bibnamefont
  {Sooryakumar}}\ and\ \bibinfo {author} {\bibfnamefont {M.~V.}\ \bibnamefont
  {Klein}},\ }\bibfield  {title} {\bibinfo {title} {Raman scattering by
  superconducting-gap excitations and their coupling to charge-density waves},\
  }\href {https://doi.org/10.1103/PhysRevLett.45.660} {\bibfield  {journal}
  {\bibinfo  {journal} {Phys. Rev. Lett.}\ }\textbf {\bibinfo {volume} {45}},\
  \bibinfo {pages} {660} (\bibinfo {year} {1980})}\BibitemShut {NoStop}%
\bibitem [{\citenamefont {Matsunaga}\ \emph {et~al.}(2014)\citenamefont
  {Matsunaga}, \citenamefont {Tsuji}, \citenamefont {Fujita}, \citenamefont
  {Sugioka}, \citenamefont {Makise}, \citenamefont {Uzawa}, \citenamefont
  {Terai}, \citenamefont {Wang}, \citenamefont {Aoki},\ and\ \citenamefont
  {Shimano}}]{Matsunaga2014-ss}%
  \BibitemOpen
  \bibfield  {author} {\bibinfo {author} {\bibfnamefont {R.}~\bibnamefont
  {Matsunaga}}, \bibinfo {author} {\bibfnamefont {N.}~\bibnamefont {Tsuji}},
  \bibinfo {author} {\bibfnamefont {H.}~\bibnamefont {Fujita}}, \bibinfo
  {author} {\bibfnamefont {A.}~\bibnamefont {Sugioka}}, \bibinfo {author}
  {\bibfnamefont {K.}~\bibnamefont {Makise}}, \bibinfo {author} {\bibfnamefont
  {Y.}~\bibnamefont {Uzawa}}, \bibinfo {author} {\bibfnamefont
  {H.}~\bibnamefont {Terai}}, \bibinfo {author} {\bibfnamefont
  {Z.}~\bibnamefont {Wang}}, \bibinfo {author} {\bibfnamefont {H.}~\bibnamefont
  {Aoki}},\ and\ \bibinfo {author} {\bibfnamefont {R.}~\bibnamefont
  {Shimano}},\ }\bibfield  {title} {\bibinfo {title} {Light-induced collective
  pseudospin precession resonating with {H}iggs mode in a superconductor},\
  }\href {https://doi.org/10.1126/science.1254697} {\bibfield  {journal}
  {\bibinfo  {journal} {Science}\ }\textbf {\bibinfo {volume} {345}},\ \bibinfo
  {pages} {1145} (\bibinfo {year} {2014})}\BibitemShut {NoStop}%
\bibitem [{\citenamefont {M\'easson}\ \emph {et~al.}(2014)\citenamefont
  {M\'easson}, \citenamefont {Gallais}, \citenamefont {Cazayous}, \citenamefont
  {Clair}, \citenamefont {Rodi\`ere}, \citenamefont {Cario},\ and\
  \citenamefont {Sacuto}}]{Measson14}%
  \BibitemOpen
  \bibfield  {author} {\bibinfo {author} {\bibfnamefont {M.-A.}\ \bibnamefont
  {M\'easson}}, \bibinfo {author} {\bibfnamefont {Y.}~\bibnamefont {Gallais}},
  \bibinfo {author} {\bibfnamefont {M.}~\bibnamefont {Cazayous}}, \bibinfo
  {author} {\bibfnamefont {B.}~\bibnamefont {Clair}}, \bibinfo {author}
  {\bibfnamefont {P.}~\bibnamefont {Rodi\`ere}}, \bibinfo {author}
  {\bibfnamefont {L.}~\bibnamefont {Cario}},\ and\ \bibinfo {author}
  {\bibfnamefont {A.}~\bibnamefont {Sacuto}},\ }\bibfield  {title} {\bibinfo
  {title} {Amplitude {H}iggs mode in the {2H-NbSe$_2$} superconductor},\ }\href
  {https://doi.org/10.1103/PhysRevB.89.060503} {\bibfield  {journal} {\bibinfo
  {journal} {Phys. Rev. B}\ }\textbf {\bibinfo {volume} {89}},\ \bibinfo
  {pages} {060503} (\bibinfo {year} {2014})}\BibitemShut {NoStop}%
\bibitem [{\citenamefont {Sherman}\ \emph {et~al.}(2015)\citenamefont
  {Sherman}, \citenamefont {Pracht}, \citenamefont {Gorshunov}, \citenamefont
  {Poran}, \citenamefont {Jesudasan}, \citenamefont {Chand}, \citenamefont
  {Raychaudhuri}, \citenamefont {Swanson}, \citenamefont {Trivedi},
  \citenamefont {Auerbach}, \citenamefont {Scheffler}, \citenamefont
  {Frydman},\ and\ \citenamefont {Dressel}}]{Sherman2015-rm}%
  \BibitemOpen
  \bibfield  {author} {\bibinfo {author} {\bibfnamefont {D.}~\bibnamefont
  {Sherman}}, \bibinfo {author} {\bibfnamefont {U.~S.}\ \bibnamefont {Pracht}},
  \bibinfo {author} {\bibfnamefont {B.}~\bibnamefont {Gorshunov}}, \bibinfo
  {author} {\bibfnamefont {S.}~\bibnamefont {Poran}}, \bibinfo {author}
  {\bibfnamefont {J.}~\bibnamefont {Jesudasan}}, \bibinfo {author}
  {\bibfnamefont {M.}~\bibnamefont {Chand}}, \bibinfo {author} {\bibfnamefont
  {P.}~\bibnamefont {Raychaudhuri}}, \bibinfo {author} {\bibfnamefont
  {M.}~\bibnamefont {Swanson}}, \bibinfo {author} {\bibfnamefont
  {N.}~\bibnamefont {Trivedi}}, \bibinfo {author} {\bibfnamefont
  {A.}~\bibnamefont {Auerbach}}, \bibinfo {author} {\bibfnamefont
  {M.}~\bibnamefont {Scheffler}}, \bibinfo {author} {\bibfnamefont
  {A.}~\bibnamefont {Frydman}},\ and\ \bibinfo {author} {\bibfnamefont
  {M.}~\bibnamefont {Dressel}},\ }\bibfield  {title} {\bibinfo {title} {The
  {H}iggs mode in disordered superconductors close to a quantum phase
  transition},\ }\href {https://doi.org/https://doi.org/10.1038/nphys3227}
  {\bibfield  {journal} {\bibinfo  {journal} {Nat. Phys.}\ }\textbf {\bibinfo
  {volume} {11}},\ \bibinfo {pages} {188} (\bibinfo {year} {2015})}\BibitemShut
  {NoStop}%
\bibitem [{\citenamefont {Bissbort}\ \emph {et~al.}(2011)\citenamefont
  {Bissbort}, \citenamefont {G{\"o}tze}, \citenamefont {Li}, \citenamefont
  {Heinze}, \citenamefont {Krauser}, \citenamefont {Weinberg}, \citenamefont
  {Becker}, \citenamefont {Sengstock},\ and\ \citenamefont
  {Hofstetter}}]{Bissbort2011-nd}%
  \BibitemOpen
  \bibfield  {author} {\bibinfo {author} {\bibfnamefont {U.}~\bibnamefont
  {Bissbort}}, \bibinfo {author} {\bibfnamefont {S.}~\bibnamefont {G{\"o}tze}},
  \bibinfo {author} {\bibfnamefont {Y.}~\bibnamefont {Li}}, \bibinfo {author}
  {\bibfnamefont {J.}~\bibnamefont {Heinze}}, \bibinfo {author} {\bibfnamefont
  {J.~S.}\ \bibnamefont {Krauser}}, \bibinfo {author} {\bibfnamefont
  {M.}~\bibnamefont {Weinberg}}, \bibinfo {author} {\bibfnamefont
  {C.}~\bibnamefont {Becker}}, \bibinfo {author} {\bibfnamefont
  {K.}~\bibnamefont {Sengstock}},\ and\ \bibinfo {author} {\bibfnamefont
  {W.}~\bibnamefont {Hofstetter}},\ }\bibfield  {title} {\bibinfo {title}
  {Detecting the amplitude mode of strongly interacting lattice bosons by
  {B}ragg scattering},\ }\href {https://doi.org/10.1103/PhysRevLett.106.205303}
  {\bibfield  {journal} {\bibinfo  {journal} {Phys. Rev. Lett.}\ }\textbf
  {\bibinfo {volume} {106}},\ \bibinfo {pages} {205303} (\bibinfo {year}
  {2011})}\BibitemShut {NoStop}%
\bibitem [{\citenamefont {Endres}\ \emph {et~al.}(2012)\citenamefont {Endres},
  \citenamefont {Fukuhara}, \citenamefont {Pekker}, \citenamefont {Cheneau},
  \citenamefont {Schauss}, \citenamefont {Gross}, \citenamefont {Demler},
  \citenamefont {Kuhr},\ and\ \citenamefont {Bloch}}]{Endres2012-ci}%
  \BibitemOpen
  \bibfield  {author} {\bibinfo {author} {\bibfnamefont {M.}~\bibnamefont
  {Endres}}, \bibinfo {author} {\bibfnamefont {T.}~\bibnamefont {Fukuhara}},
  \bibinfo {author} {\bibfnamefont {D.}~\bibnamefont {Pekker}}, \bibinfo
  {author} {\bibfnamefont {M.}~\bibnamefont {Cheneau}}, \bibinfo {author}
  {\bibfnamefont {P.}~\bibnamefont {Schauss}}, \bibinfo {author} {\bibfnamefont
  {C.}~\bibnamefont {Gross}}, \bibinfo {author} {\bibfnamefont
  {E.}~\bibnamefont {Demler}}, \bibinfo {author} {\bibfnamefont
  {S.}~\bibnamefont {Kuhr}},\ and\ \bibinfo {author} {\bibfnamefont
  {I.}~\bibnamefont {Bloch}},\ }\bibfield  {title} {\bibinfo {title} {The
  `{H}iggs' amplitude mode at the two-dimensional superfluid/{M}ott insulator
  transition},\ }\href {https://doi.org/10.1038/nature11255} {\bibfield
  {journal} {\bibinfo  {journal} {Nature}\ }\textbf {\bibinfo {volume} {487}},\
  \bibinfo {pages} {454} (\bibinfo {year} {2012})}\BibitemShut {NoStop}%
\bibitem [{\citenamefont {L{\'e}onard}\ \emph {et~al.}(2017)\citenamefont
  {L{\'e}onard}, \citenamefont {Morales}, \citenamefont {Zupancic},
  \citenamefont {Donner},\ and\ \citenamefont {Esslinger}}]{Leonard2017-sq}%
  \BibitemOpen
  \bibfield  {author} {\bibinfo {author} {\bibfnamefont {J.}~\bibnamefont
  {L{\'e}onard}}, \bibinfo {author} {\bibfnamefont {A.}~\bibnamefont
  {Morales}}, \bibinfo {author} {\bibfnamefont {P.}~\bibnamefont {Zupancic}},
  \bibinfo {author} {\bibfnamefont {T.}~\bibnamefont {Donner}},\ and\ \bibinfo
  {author} {\bibfnamefont {T.}~\bibnamefont {Esslinger}},\ }\bibfield  {title}
  {\bibinfo {title} {Monitoring and manipulating {H}iggs and {G}oldstone modes
  in a supersolid quantum gas},\ }\href
  {https://doi.org/10.1126/science.aan2608} {\bibfield  {journal} {\bibinfo
  {journal} {Science}\ }\textbf {\bibinfo {volume} {358}},\ \bibinfo {pages}
  {1415} (\bibinfo {year} {2017})}\BibitemShut {NoStop}%
\bibitem [{\citenamefont {Behrle}\ \emph {et~al.}(2018)\citenamefont {Behrle},
  \citenamefont {Harrison}, \citenamefont {Kombe}, \citenamefont {Gao},
  \citenamefont {Link}, \citenamefont {Bernier}, \citenamefont {Kollath},\ and\
  \citenamefont {K{\"o}hl}}]{Behrle2018-pn}%
  \BibitemOpen
  \bibfield  {author} {\bibinfo {author} {\bibfnamefont {A.}~\bibnamefont
  {Behrle}}, \bibinfo {author} {\bibfnamefont {T.}~\bibnamefont {Harrison}},
  \bibinfo {author} {\bibfnamefont {J.}~\bibnamefont {Kombe}}, \bibinfo
  {author} {\bibfnamefont {K.}~\bibnamefont {Gao}}, \bibinfo {author}
  {\bibfnamefont {M.}~\bibnamefont {Link}}, \bibinfo {author} {\bibfnamefont
  {J.-S.}\ \bibnamefont {Bernier}}, \bibinfo {author} {\bibfnamefont
  {C.}~\bibnamefont {Kollath}},\ and\ \bibinfo {author} {\bibfnamefont
  {M.}~\bibnamefont {K{\"o}hl}},\ }\bibfield  {title} {\bibinfo {title}
  {{H}iggs mode in a strongly interacting fermionic superfluid},\ }\href
  {https://doi.org/10.1038/s41567-018-0128-6} {\bibfield  {journal} {\bibinfo
  {journal} {Nat. Phys.}\ }\textbf {\bibinfo {volume} {14}},\ \bibinfo {pages}
  {781} (\bibinfo {year} {2018})}\BibitemShut {NoStop}%
\bibitem [{\citenamefont {Dyke}\ \emph {et~al.}(2024)\citenamefont {Dyke},
  \citenamefont {Musolino}, \citenamefont {Kurkjian}, \citenamefont
  {Ahmed-Braun}, \citenamefont {Pennings}, \citenamefont {Herrera},
  \citenamefont {Hoinka}, \citenamefont {Kokkelmans}, \citenamefont {Colussi},\
  and\ \citenamefont {Vale}}]{Dyke2024-kx}%
  \BibitemOpen
  \bibfield  {author} {\bibinfo {author} {\bibfnamefont {P.}~\bibnamefont
  {Dyke}}, \bibinfo {author} {\bibfnamefont {S.}~\bibnamefont {Musolino}},
  \bibinfo {author} {\bibfnamefont {H.}~\bibnamefont {Kurkjian}}, \bibinfo
  {author} {\bibfnamefont {D.~J.~M.}\ \bibnamefont {Ahmed-Braun}}, \bibinfo
  {author} {\bibfnamefont {A.}~\bibnamefont {Pennings}}, \bibinfo {author}
  {\bibfnamefont {I.}~\bibnamefont {Herrera}}, \bibinfo {author} {\bibfnamefont
  {S.}~\bibnamefont {Hoinka}}, \bibinfo {author} {\bibfnamefont {S.~J. J.
  M.~F.}\ \bibnamefont {Kokkelmans}}, \bibinfo {author} {\bibfnamefont {V.~E.}\
  \bibnamefont {Colussi}},\ and\ \bibinfo {author} {\bibfnamefont {C.~J.}\
  \bibnamefont {Vale}},\ }\bibfield  {title} {\bibinfo {title} {{H}iggs
  oscillations in a unitary {F}ermi superfluid},\ }\href
  {https://doi.org/10.1103/PhysRevLett.132.223402} {\bibfield  {journal}
  {\bibinfo  {journal} {Phys. Rev. Lett.}\ }\textbf {\bibinfo {volume} {132}},\
  \bibinfo {pages} {223402} (\bibinfo {year} {2024})}\BibitemShut {NoStop}%
\bibitem [{\citenamefont {Kell}\ \emph {et~al.}(2024)\citenamefont {Kell},
  \citenamefont {Breyer}, \citenamefont {Eberz},\ and\ \citenamefont
  {Köhl}}]{Kell2024-jx}%
  \BibitemOpen
  \bibfield  {author} {\bibinfo {author} {\bibfnamefont {A.}~\bibnamefont
  {Kell}}, \bibinfo {author} {\bibfnamefont {M.}~\bibnamefont {Breyer}},
  \bibinfo {author} {\bibfnamefont {D.}~\bibnamefont {Eberz}},\ and\ \bibinfo
  {author} {\bibfnamefont {M.}~\bibnamefont {Köhl}},\ }\bibfield  {title}
  {\bibinfo {title} {Exciting the higgs mode in a strongly interacting fermi
  gas by interaction modulation},\ }\href
  {https://doi.org/10.1103/PhysRevLett.133.150403} {\bibfield  {journal}
  {\bibinfo  {journal} {Phys. Rev. Lett.}\ }\textbf {\bibinfo {volume} {133}},\
  \bibinfo {pages} {150403} (\bibinfo {year} {2024})}\BibitemShut {NoStop}%
\bibitem [{\citenamefont {Bayha}\ \emph {et~al.}(2020)\citenamefont {Bayha},
  \citenamefont {Holten}, \citenamefont {Klemt}, \citenamefont {Subramanian},
  \citenamefont {Bjerlin}, \citenamefont {Reimann}, \citenamefont {Bruun},
  \citenamefont {Preiss},\ and\ \citenamefont {Jochim}}]{Bayha2020-ek}%
  \BibitemOpen
  \bibfield  {author} {\bibinfo {author} {\bibfnamefont {L.}~\bibnamefont
  {Bayha}}, \bibinfo {author} {\bibfnamefont {M.}~\bibnamefont {Holten}},
  \bibinfo {author} {\bibfnamefont {R.}~\bibnamefont {Klemt}}, \bibinfo
  {author} {\bibfnamefont {K.}~\bibnamefont {Subramanian}}, \bibinfo {author}
  {\bibfnamefont {J.}~\bibnamefont {Bjerlin}}, \bibinfo {author} {\bibfnamefont
  {S.~M.}\ \bibnamefont {Reimann}}, \bibinfo {author} {\bibfnamefont {G.~M.}\
  \bibnamefont {Bruun}}, \bibinfo {author} {\bibfnamefont {P.~M.}\ \bibnamefont
  {Preiss}},\ and\ \bibinfo {author} {\bibfnamefont {S.}~\bibnamefont
  {Jochim}},\ }\bibfield  {title} {\bibinfo {title} {Observing the emergence of
  a quantum phase transition shell by shell},\ }\href
  {https://doi.org/https://doi.org/10.1038/s41586-020-2936-y} {\bibfield
  {journal} {\bibinfo  {journal} {Nature}\ }\textbf {\bibinfo {volume} {587}},\
  \bibinfo {pages} {583} (\bibinfo {year} {2020})}\BibitemShut {NoStop}%
\bibitem [{\citenamefont {Skulte}\ \emph {et~al.}(2021)\citenamefont {Skulte},
  \citenamefont {Broers}, \citenamefont {Cosme},\ and\ \citenamefont
  {Mathey}}]{Skulte2021}%
  \BibitemOpen
  \bibfield  {author} {\bibinfo {author} {\bibfnamefont {J.}~\bibnamefont
  {Skulte}}, \bibinfo {author} {\bibfnamefont {L.}~\bibnamefont {Broers}},
  \bibinfo {author} {\bibfnamefont {J.~G.}\ \bibnamefont {Cosme}},\ and\
  \bibinfo {author} {\bibfnamefont {L.}~\bibnamefont {Mathey}},\ }\bibfield
  {title} {\bibinfo {title} {Vortex and soliton dynamics in
  particle-hole-symmetric superfluids},\ }\href
  {https://doi.org/10.1103/PhysRevResearch.3.043109} {\bibfield  {journal}
  {\bibinfo  {journal} {Phys. Rev. Res.}\ }\textbf {\bibinfo {volume} {3}},\
  \bibinfo {pages} {043109} (\bibinfo {year} {2021})}\BibitemShut {NoStop}%
\bibitem [{sup()}]{supmat}%
  \BibitemOpen
  \href@noop {} {}\bibinfo {note} {See Supplemental Material.}\BibitemShut
  {Stop}%
\bibitem [{\citenamefont {Hueck}\ \emph {et~al.}(2018)\citenamefont {Hueck},
  \citenamefont {Luick}, \citenamefont {Sobirey}, \citenamefont {Siegl},
  \citenamefont {Lompe},\ and\ \citenamefont {Moritz}}]{Hueck2018-cs}%
  \BibitemOpen
  \bibfield  {author} {\bibinfo {author} {\bibfnamefont {K.}~\bibnamefont
  {Hueck}}, \bibinfo {author} {\bibfnamefont {N.}~\bibnamefont {Luick}},
  \bibinfo {author} {\bibfnamefont {L.}~\bibnamefont {Sobirey}}, \bibinfo
  {author} {\bibfnamefont {J.}~\bibnamefont {Siegl}}, \bibinfo {author}
  {\bibfnamefont {T.}~\bibnamefont {Lompe}},\ and\ \bibinfo {author}
  {\bibfnamefont {H.}~\bibnamefont {Moritz}},\ }\bibfield  {title} {\bibinfo
  {title} {Two-dimensional homogeneous {F}ermi gases},\ }\href
  {https://doi.org/10.1103/PhysRevLett.120.060402} {\bibfield  {journal}
  {\bibinfo  {journal} {Phys. Rev. Lett.}\ }\textbf {\bibinfo {volume} {120}},\
  \bibinfo {pages} {060402} (\bibinfo {year} {2018})}\BibitemShut {NoStop}%
\bibitem [{\citenamefont {Petrov}\ and\ \citenamefont
  {Shlyapnikov}(2001)}]{Petrov2001-kn}%
  \BibitemOpen
  \bibfield  {author} {\bibinfo {author} {\bibfnamefont {D.~S.}\ \bibnamefont
  {Petrov}}\ and\ \bibinfo {author} {\bibfnamefont {G.~V.}\ \bibnamefont
  {Shlyapnikov}},\ }\bibfield  {title} {\bibinfo {title} {Interatomic
  collisions in a tightly confined {B}ose gas},\ }\href
  {https://doi.org/10.1103/PhysRevA.64.012706} {\bibfield  {journal} {\bibinfo
  {journal} {Phys. Rev. A}\ }\textbf {\bibinfo {volume} {64}},\ \bibinfo
  {pages} {012706} (\bibinfo {year} {2001})}\BibitemShut {NoStop}%
\bibitem [{\citenamefont {Levinsen}\ and\ \citenamefont
  {Parish}(2015)}]{Levinsen2015-du}%
  \BibitemOpen
  \bibfield  {author} {\bibinfo {author} {\bibfnamefont {J.}~\bibnamefont
  {Levinsen}}\ and\ \bibinfo {author} {\bibfnamefont {M.~M.}\ \bibnamefont
  {Parish}},\ }\bibfield  {title} {\bibinfo {title} {Strongly interacting
  two-dimensional {F}ermi gases},\ }\href
  {https://doi.org/10.1142/9789814667746\_0001} {\ \bibinfo {series} {Annual
  Review of Cold Atoms and Molecules},\ \textbf {\bibinfo {volume} {3}},\
  \bibinfo {pages} {1} (\bibinfo {year} {2015})}\BibitemShut {NoStop}%
\bibitem [{\citenamefont {Turlapov}\ and\ \citenamefont
  {Yu~Kagan}(2017)}]{Turlapov2017-nl}%
  \BibitemOpen
  \bibfield  {author} {\bibinfo {author} {\bibfnamefont {A.~V.}\ \bibnamefont
  {Turlapov}}\ and\ \bibinfo {author} {\bibfnamefont {M.}~\bibnamefont
  {Yu~Kagan}},\ }\bibfield  {title} {\bibinfo {title} {{{F}ermi-to-Bose}
  crossover in a trapped quasi-2d gas of fermionic atoms},\ }\href
  {https://doi.org/10.1088/1361-648X/aa7ad9} {\bibfield  {journal} {\bibinfo
  {journal} {J. Phys. Condens. Matter}\ }\textbf {\bibinfo {volume} {29}},\
  \bibinfo {pages} {383004} (\bibinfo {year} {2017})}\BibitemShut {NoStop}%
\bibitem [{\citenamefont {Veeravalli}\ \emph {et~al.}(2008)\citenamefont
  {Veeravalli}, \citenamefont {Kuhnle}, \citenamefont {Dyke},\ and\
  \citenamefont {Vale}}]{Veeravalli2008-hl}%
  \BibitemOpen
  \bibfield  {author} {\bibinfo {author} {\bibfnamefont {G.}~\bibnamefont
  {Veeravalli}}, \bibinfo {author} {\bibfnamefont {E.}~\bibnamefont {Kuhnle}},
  \bibinfo {author} {\bibfnamefont {P.}~\bibnamefont {Dyke}},\ and\ \bibinfo
  {author} {\bibfnamefont {C.~J.}\ \bibnamefont {Vale}},\ }\bibfield  {title}
  {\bibinfo {title} {{B}ragg spectroscopy of a strongly interacting {F}ermi
  gas},\ }\href {https://doi.org/10.1103/PhysRevLett.101.250403} {\bibfield
  {journal} {\bibinfo  {journal} {Phys. Rev. Lett.}\ }\textbf {\bibinfo
  {volume} {101}},\ \bibinfo {pages} {250403} (\bibinfo {year}
  {2008})}\BibitemShut {NoStop}%
\bibitem [{\citenamefont {Hoinka}\ \emph {et~al.}(2017)\citenamefont {Hoinka},
  \citenamefont {Dyke}, \citenamefont {Lingham}, \citenamefont {Kinnunen},
  \citenamefont {Bruun},\ and\ \citenamefont {Vale}}]{Hoinka2017-qp}%
  \BibitemOpen
  \bibfield  {author} {\bibinfo {author} {\bibfnamefont {S.}~\bibnamefont
  {Hoinka}}, \bibinfo {author} {\bibfnamefont {P.}~\bibnamefont {Dyke}},
  \bibinfo {author} {\bibfnamefont {M.~G.}\ \bibnamefont {Lingham}}, \bibinfo
  {author} {\bibfnamefont {J.~J.}\ \bibnamefont {Kinnunen}}, \bibinfo {author}
  {\bibfnamefont {G.~M.}\ \bibnamefont {Bruun}},\ and\ \bibinfo {author}
  {\bibfnamefont {C.~J.}\ \bibnamefont {Vale}},\ }\bibfield  {title} {\bibinfo
  {title} {{G}oldstone mode and pair-breaking excitations in atomic {F}ermi
  superfluids},\ }\href {https://doi.org/10.1038/nphys4187} {\bibfield
  {journal} {\bibinfo  {journal} {Nat. Phys.}\ }\textbf {\bibinfo {volume}
  {13}},\ \bibinfo {pages} {943} (\bibinfo {year} {2017})}\BibitemShut
  {NoStop}%
\bibitem [{\citenamefont {Sobirey}\ \emph {et~al.}(2021)\citenamefont
  {Sobirey}, \citenamefont {Luick}, \citenamefont {Bohlen}, \citenamefont
  {Biss}, \citenamefont {Moritz},\ and\ \citenamefont
  {Lompe}}]{Sobirey2021-xy}%
  \BibitemOpen
  \bibfield  {author} {\bibinfo {author} {\bibfnamefont {L.}~\bibnamefont
  {Sobirey}}, \bibinfo {author} {\bibfnamefont {N.}~\bibnamefont {Luick}},
  \bibinfo {author} {\bibfnamefont {M.}~\bibnamefont {Bohlen}}, \bibinfo
  {author} {\bibfnamefont {H.}~\bibnamefont {Biss}}, \bibinfo {author}
  {\bibfnamefont {H.}~\bibnamefont {Moritz}},\ and\ \bibinfo {author}
  {\bibfnamefont {T.}~\bibnamefont {Lompe}},\ }\href
  {https://doi.org/10.1126/science.abc8793} {\bibinfo {title} {Observation of
  superfluidity in a strongly correlated two-dimensional {F}ermi gas}}
  (\bibinfo {year} {2021})\BibitemShut {NoStop}%
\bibitem [{\citenamefont {Biss}\ \emph {et~al.}(2022)\citenamefont {Biss},
  \citenamefont {Sobirey}, \citenamefont {Luick}, \citenamefont {Bohlen},
  \citenamefont {Kinnunen}, \citenamefont {Bruun}, \citenamefont {Lompe},\ and\
  \citenamefont {Moritz}}]{Biss2022-ra}%
  \BibitemOpen
  \bibfield  {author} {\bibinfo {author} {\bibfnamefont {H.}~\bibnamefont
  {Biss}}, \bibinfo {author} {\bibfnamefont {L.}~\bibnamefont {Sobirey}},
  \bibinfo {author} {\bibfnamefont {N.}~\bibnamefont {Luick}}, \bibinfo
  {author} {\bibfnamefont {M.}~\bibnamefont {Bohlen}}, \bibinfo {author}
  {\bibfnamefont {J.~J.}\ \bibnamefont {Kinnunen}}, \bibinfo {author}
  {\bibfnamefont {G.~M.}\ \bibnamefont {Bruun}}, \bibinfo {author}
  {\bibfnamefont {T.}~\bibnamefont {Lompe}},\ and\ \bibinfo {author}
  {\bibfnamefont {H.}~\bibnamefont {Moritz}},\ }\bibfield  {title} {\bibinfo
  {title} {Excitation spectrum and superfluid gap of an ultracold {F}ermi
  gas},\ }\href {https://doi.org/10.1103/PhysRevLett.128.100401} {\bibfield
  {journal} {\bibinfo  {journal} {Phys. Rev. Lett.}\ }\textbf {\bibinfo
  {volume} {128}},\ \bibinfo {pages} {100401} (\bibinfo {year}
  {2022})}\BibitemShut {NoStop}%
\bibitem [{\citenamefont {Salasnich}(2007)}]{Salasnich2007}%
  \BibitemOpen
  \bibfield  {author} {\bibinfo {author} {\bibfnamefont {L.}~\bibnamefont
  {Salasnich}},\ }\bibfield  {title} {\bibinfo {title} {Condensate fraction of
  a two-dimensional attractive {F}ermi gas},\ }\href
  {https://doi.org/10.1103/PhysRevA.76.015601} {\bibfield  {journal} {\bibinfo
  {journal} {Phys. Rev. A}\ }\textbf {\bibinfo {volume} {76}},\ \bibinfo
  {pages} {015601} (\bibinfo {year} {2007})}\BibitemShut {NoStop}%
\bibitem [{\citenamefont {Shi}\ \emph {et~al.}(2015)\citenamefont {Shi},
  \citenamefont {Chiesa},\ and\ \citenamefont {Zhang}}]{Shi2015}%
  \BibitemOpen
  \bibfield  {author} {\bibinfo {author} {\bibfnamefont {H.}~\bibnamefont
  {Shi}}, \bibinfo {author} {\bibfnamefont {S.}~\bibnamefont {Chiesa}},\ and\
  \bibinfo {author} {\bibfnamefont {S.}~\bibnamefont {Zhang}},\ }\bibfield
  {title} {\bibinfo {title} {Ground-state properties of strongly interacting
  {F}ermi gases in two dimensions},\ }\href
  {https://doi.org/10.1103/PhysRevA.92.033603} {\bibfield  {journal} {\bibinfo
  {journal} {Phys. Rev. A}\ }\textbf {\bibinfo {volume} {92}},\ \bibinfo
  {pages} {033603} (\bibinfo {year} {2015})}\BibitemShut {NoStop}%
\bibitem [{\citenamefont {Verresen}\ \emph {et~al.}(2019)\citenamefont
  {Verresen}, \citenamefont {Moessner},\ and\ \citenamefont
  {Pollmann}}]{Verresen2019-jp}%
  \BibitemOpen
  \bibfield  {author} {\bibinfo {author} {\bibfnamefont {R.}~\bibnamefont
  {Verresen}}, \bibinfo {author} {\bibfnamefont {R.}~\bibnamefont {Moessner}},\
  and\ \bibinfo {author} {\bibfnamefont {F.}~\bibnamefont {Pollmann}},\
  }\bibfield  {title} {\bibinfo {title} {Avoided quasiparticle decay from
  strong quantum interactions},\ }\href
  {https://doi.org/10.1038/s41567-019-0535-3} {\bibfield  {journal} {\bibinfo
  {journal} {Nat. Phys.}\ }\textbf {\bibinfo {volume} {15}},\ \bibinfo {pages}
  {750} (\bibinfo {year} {2019})}\BibitemShut {NoStop}%
\bibitem [{\citenamefont {Lorenzana}\ and\ \citenamefont
  {Seibold}(2024)}]{Lorenzana24}%
  \BibitemOpen
  \bibfield  {author} {\bibinfo {author} {\bibfnamefont {J.}~\bibnamefont
  {Lorenzana}}\ and\ \bibinfo {author} {\bibfnamefont {G.}~\bibnamefont
  {Seibold}},\ }\bibfield  {title} {\bibinfo {title} {Long-lived {H}iggs modes
  in strongly correlated condensates},\ }\href
  {https://doi.org/10.1103/PhysRevLett.132.026501} {\bibfield  {journal}
  {\bibinfo  {journal} {Phys. Rev. Lett.}\ }\textbf {\bibinfo {volume} {132}},\
  \bibinfo {pages} {026501} (\bibinfo {year} {2024})}\BibitemShut {NoStop}%
\bibitem [{\citenamefont {Cea}\ \emph {et~al.}(2015)\citenamefont {Cea},
  \citenamefont {Castellani}, \citenamefont {Seibold},\ and\ \citenamefont
  {Benfatto}}]{Benfatto15}%
  \BibitemOpen
  \bibfield  {author} {\bibinfo {author} {\bibfnamefont {T.}~\bibnamefont
  {Cea}}, \bibinfo {author} {\bibfnamefont {C.}~\bibnamefont {Castellani}},
  \bibinfo {author} {\bibfnamefont {G.}~\bibnamefont {Seibold}},\ and\ \bibinfo
  {author} {\bibfnamefont {L.}~\bibnamefont {Benfatto}},\ }\bibfield  {title}
  {\bibinfo {title} {Nonrelativistic dynamics of the amplitude ({H}iggs) mode
  in superconductors},\ }\href {https://doi.org/10.1103/PhysRevLett.115.157002}
  {\bibfield  {journal} {\bibinfo  {journal} {Phys. Rev. Lett.}\ }\textbf
  {\bibinfo {volume} {115}},\ \bibinfo {pages} {157002} (\bibinfo {year}
  {2015})}\BibitemShut {NoStop}%
\bibitem [{\citenamefont {Cea}\ \emph {et~al.}(2016)\citenamefont {Cea},
  \citenamefont {Castellani},\ and\ \citenamefont {Benfatto}}]{Benfatto16}%
  \BibitemOpen
  \bibfield  {author} {\bibinfo {author} {\bibfnamefont {T.}~\bibnamefont
  {Cea}}, \bibinfo {author} {\bibfnamefont {C.}~\bibnamefont {Castellani}},\
  and\ \bibinfo {author} {\bibfnamefont {L.}~\bibnamefont {Benfatto}},\
  }\bibfield  {title} {\bibinfo {title} {Nonlinear optical effects and
  third-harmonic generation in superconductors: Cooper pairs versus {H}iggs
  mode contribution},\ }\href {https://doi.org/10.1103/PhysRevB.93.180507}
  {\bibfield  {journal} {\bibinfo  {journal} {Phys. Rev. B}\ }\textbf {\bibinfo
  {volume} {93}},\ \bibinfo {pages} {180507} (\bibinfo {year}
  {2016})}\BibitemShut {NoStop}%
\bibitem [{\citenamefont {Kurkjian}\ \emph {et~al.}(2019)\citenamefont
  {Kurkjian}, \citenamefont {Klimin}, \citenamefont {Tempere},\ and\
  \citenamefont {Castin}}]{Kurkjian2019-ng}%
  \BibitemOpen
  \bibfield  {author} {\bibinfo {author} {\bibfnamefont {H.}~\bibnamefont
  {Kurkjian}}, \bibinfo {author} {\bibfnamefont {S.~N.}\ \bibnamefont
  {Klimin}}, \bibinfo {author} {\bibfnamefont {J.}~\bibnamefont {Tempere}},\
  and\ \bibinfo {author} {\bibfnamefont {Y.}~\bibnamefont {Castin}},\
  }\bibfield  {title} {\bibinfo {title} {Pair-breaking collective branch in
  {BCS} superconductors and superfluid {F}ermi gases},\ }\href
  {https://doi.org/10.1103/PhysRevLett.122.093403} {\bibfield  {journal}
  {\bibinfo  {journal} {Phys. Rev. Lett.}\ }\textbf {\bibinfo {volume} {122}},\
  \bibinfo {pages} {093403} (\bibinfo {year} {2019})}\BibitemShut {NoStop}%
\bibitem [{\citenamefont {Kurkjian}\ \emph {et~al.}(2020)\citenamefont
  {Kurkjian}, \citenamefont {Tempere},\ and\ \citenamefont
  {Klimin}}]{Kurkjian2020-xt}%
  \BibitemOpen
  \bibfield  {author} {\bibinfo {author} {\bibfnamefont {H.}~\bibnamefont
  {Kurkjian}}, \bibinfo {author} {\bibfnamefont {J.}~\bibnamefont {Tempere}},\
  and\ \bibinfo {author} {\bibfnamefont {S.~N.}\ \bibnamefont {Klimin}},\
  }\bibfield  {title} {\bibinfo {title} {Linear response of a superfluid fermi
  gas inside its pair-breaking continuum},\ }\href
  {https://doi.org/10.1038/s41598-020-65371-9} {\bibfield  {journal} {\bibinfo
  {journal} {Sci. Rep.}\ }\textbf {\bibinfo {volume} {10}},\ \bibinfo {pages}
  {11591} (\bibinfo {year} {2020})}\BibitemShut {NoStop}%
\bibitem [{\citenamefont {Vaswani}\ \emph {et~al.}(2021)\citenamefont
  {Vaswani}, \citenamefont {Kang}, \citenamefont {Mootz}, \citenamefont {Luo},
  \citenamefont {Yang}, \citenamefont {Sundahl}, \citenamefont {Cheng},
  \citenamefont {Huang}, \citenamefont {Kim}, \citenamefont {Liu},
  \citenamefont {Collantes}, \citenamefont {Hellstrom}, \citenamefont
  {Perakis}, \citenamefont {Eom},\ and\ \citenamefont {Wang}}]{Vaswani2021-mw}%
  \BibitemOpen
  \bibfield  {author} {\bibinfo {author} {\bibfnamefont {C.}~\bibnamefont
  {Vaswani}}, \bibinfo {author} {\bibfnamefont {J.~H.}\ \bibnamefont {Kang}},
  \bibinfo {author} {\bibfnamefont {M.}~\bibnamefont {Mootz}}, \bibinfo
  {author} {\bibfnamefont {L.}~\bibnamefont {Luo}}, \bibinfo {author}
  {\bibfnamefont {X.}~\bibnamefont {Yang}}, \bibinfo {author} {\bibfnamefont
  {C.}~\bibnamefont {Sundahl}}, \bibinfo {author} {\bibfnamefont
  {D.}~\bibnamefont {Cheng}}, \bibinfo {author} {\bibfnamefont
  {C.}~\bibnamefont {Huang}}, \bibinfo {author} {\bibfnamefont {R.~H.~J.}\
  \bibnamefont {Kim}}, \bibinfo {author} {\bibfnamefont {Z.}~\bibnamefont
  {Liu}}, \bibinfo {author} {\bibfnamefont {Y.~G.}\ \bibnamefont {Collantes}},
  \bibinfo {author} {\bibfnamefont {E.~E.}\ \bibnamefont {Hellstrom}}, \bibinfo
  {author} {\bibfnamefont {I.~E.}\ \bibnamefont {Perakis}}, \bibinfo {author}
  {\bibfnamefont {C.~B.}\ \bibnamefont {Eom}},\ and\ \bibinfo {author}
  {\bibfnamefont {J.}~\bibnamefont {Wang}},\ }\bibfield  {title} {\bibinfo
  {title} {Light quantum control of persisting {H}iggs modes in iron-based
  superconductors},\ }\href {https://doi.org/10.1038/s41467-020-20350-6}
  {\bibfield  {journal} {\bibinfo  {journal} {Nat. Commun.}\ }\textbf {\bibinfo
  {volume} {12}},\ \bibinfo {pages} {258} (\bibinfo {year} {2021})}\BibitemShut
  {NoStop}%
\bibitem [{\citenamefont {Hannibal}\ \emph {et~al.}(2015)\citenamefont
  {Hannibal}, \citenamefont {Kettmann}, \citenamefont {Croitoru}, \citenamefont
  {Vagov}, \citenamefont {Axt},\ and\ \citenamefont {Kuhn}}]{Hannibal2015}%
  \BibitemOpen
  \bibfield  {author} {\bibinfo {author} {\bibfnamefont {S.}~\bibnamefont
  {Hannibal}}, \bibinfo {author} {\bibfnamefont {P.}~\bibnamefont {Kettmann}},
  \bibinfo {author} {\bibfnamefont {M.~D.}\ \bibnamefont {Croitoru}}, \bibinfo
  {author} {\bibfnamefont {A.}~\bibnamefont {Vagov}}, \bibinfo {author}
  {\bibfnamefont {V.~M.}\ \bibnamefont {Axt}},\ and\ \bibinfo {author}
  {\bibfnamefont {T.}~\bibnamefont {Kuhn}},\ }\bibfield  {title} {\bibinfo
  {title} {Quench dynamics of an ultracold {F}ermi gas in the {BCS} regime:
  Spectral properties and confinement-induced breakdown of the {H}iggs mode},\
  }\href {https://doi.org/10.1103/PhysRevA.91.043630} {\bibfield  {journal}
  {\bibinfo  {journal} {Phys. Rev. A}\ }\textbf {\bibinfo {volume} {91}},\
  \bibinfo {pages} {043630} (\bibinfo {year} {2015})}\BibitemShut {NoStop}%
\bibitem [{\citenamefont {Hannibal}\ \emph {et~al.}(2018)\citenamefont
  {Hannibal}, \citenamefont {Kettmann}, \citenamefont {Croitoru}, \citenamefont
  {Axt},\ and\ \citenamefont {Kuhn}}]{Hannibal2018}%
  \BibitemOpen
  \bibfield  {author} {\bibinfo {author} {\bibfnamefont {S.}~\bibnamefont
  {Hannibal}}, \bibinfo {author} {\bibfnamefont {P.}~\bibnamefont {Kettmann}},
  \bibinfo {author} {\bibfnamefont {M.~D.}\ \bibnamefont {Croitoru}}, \bibinfo
  {author} {\bibfnamefont {V.~M.}\ \bibnamefont {Axt}},\ and\ \bibinfo {author}
  {\bibfnamefont {T.}~\bibnamefont {Kuhn}},\ }\bibfield  {title} {\bibinfo
  {title} {Dynamical vanishing of the order parameter in a confined
  bardeen-cooper-schrieffer {F}ermi gas after an interaction quench},\ }\href
  {https://doi.org/10.1103/PhysRevA.97.013619} {\bibfield  {journal} {\bibinfo
  {journal} {Phys. Rev. A}\ }\textbf {\bibinfo {volume} {97}},\ \bibinfo
  {pages} {013619} (\bibinfo {year} {2018})}\BibitemShut {NoStop}%
\bibitem [{\citenamefont {Collado}\ \emph {et~al.}(2023)\citenamefont
  {Collado}, \citenamefont {Defenu},\ and\ \citenamefont {Lorenzana}}]{HP2023}%
  \BibitemOpen
  \bibfield  {author} {\bibinfo {author} {\bibfnamefont {H.~P.~O.}\
  \bibnamefont {Collado}}, \bibinfo {author} {\bibfnamefont {N.}~\bibnamefont
  {Defenu}},\ and\ \bibinfo {author} {\bibfnamefont {J.}~\bibnamefont
  {Lorenzana}},\ }\bibfield  {title} {\bibinfo {title} {Engineering {H}iggs
  dynamics by spectral singularities},\ }\href
  {https://doi.org/10.1103/PhysRevResearch.5.023011} {\bibfield  {journal}
  {\bibinfo  {journal} {Phys. Rev. Res.}\ }\textbf {\bibinfo {volume} {5}},\
  \bibinfo {pages} {023011} (\bibinfo {year} {2023})}\BibitemShut {NoStop}%
\bibitem [{\citenamefont {Nakayama}\ \emph {et~al.}(2015)\citenamefont
  {Nakayama}, \citenamefont {Danshita}, \citenamefont {Nikuni},\ and\
  \citenamefont {Tsuchiya}}]{Nakayama2015-ro}%
  \BibitemOpen
  \bibfield  {author} {\bibinfo {author} {\bibfnamefont {T.}~\bibnamefont
  {Nakayama}}, \bibinfo {author} {\bibfnamefont {I.}~\bibnamefont {Danshita}},
  \bibinfo {author} {\bibfnamefont {T.}~\bibnamefont {Nikuni}},\ and\ \bibinfo
  {author} {\bibfnamefont {S.}~\bibnamefont {Tsuchiya}},\ }\bibfield  {title}
  {\bibinfo {title} {Fano resonance through {H}iggs bound states in tunneling
  of {{N}ambu-{G}oldstone} modes},\ }\href
  {https://doi.org/10.1103/PhysRevA.92.043610} {\bibfield  {journal} {\bibinfo
  {journal} {Phys. Rev. A}\ }\textbf {\bibinfo {volume} {92}},\ \bibinfo
  {pages} {043610} (\bibinfo {year} {2015})}\BibitemShut {NoStop}%
\bibitem [{\citenamefont {Nakayama}\ and\ \citenamefont
  {Tsuchiya}(2019)}]{Nakayama2019-xy}%
  \BibitemOpen
  \bibfield  {author} {\bibinfo {author} {\bibfnamefont {T.}~\bibnamefont
  {Nakayama}}\ and\ \bibinfo {author} {\bibfnamefont {S.}~\bibnamefont
  {Tsuchiya}},\ }\bibfield  {title} {\bibinfo {title} {Perfect transmission of
  {H}iggs modes via antibound states},\ }\href
  {https://doi.org/10.1103/PhysRevA.100.063612} {\bibfield  {journal} {\bibinfo
   {journal} {Phys. Rev. A}\ }\textbf {\bibinfo {volume} {100}},\ \bibinfo
  {pages} {063612} (\bibinfo {year} {2019})}\BibitemShut {NoStop}%
\bibitem [{\citenamefont {Leggett}(1966)}]{Legget66}%
  \BibitemOpen
  \bibfield  {author} {\bibinfo {author} {\bibfnamefont {A.~J.}\ \bibnamefont
  {Leggett}},\ }\bibfield  {title} {\bibinfo {title} {Number-phase fluctuations
  in two-band superconductors},\ }\href {https://doi.org/10.1143/PTP.36.901}
  {\bibfield  {journal} {\bibinfo  {journal} {Progress of Theoretical Physics}\
  }\textbf {\bibinfo {volume} {36}},\ \bibinfo {pages} {901} (\bibinfo {year}
  {1966})}\BibitemShut {NoStop}%
\bibitem [{\citenamefont {Krull}\ \emph {et~al.}(2016)\citenamefont {Krull},
  \citenamefont {Bittner}, \citenamefont {Uhrig}, \citenamefont {Manske},\ and\
  \citenamefont {Schnyder}}]{Krull16}%
  \BibitemOpen
  \bibfield  {author} {\bibinfo {author} {\bibfnamefont {H.}~\bibnamefont
  {Krull}}, \bibinfo {author} {\bibfnamefont {N.}~\bibnamefont {Bittner}},
  \bibinfo {author} {\bibfnamefont {G.~S.}\ \bibnamefont {Uhrig}}, \bibinfo
  {author} {\bibfnamefont {D.}~\bibnamefont {Manske}},\ and\ \bibinfo {author}
  {\bibfnamefont {A.~P.}\ \bibnamefont {Schnyder}},\ }\bibfield  {title}
  {\bibinfo {title} {Coupling of {H}iggs and {L}eggett modes in non-equilibrium
  superconductors},\ }\href {https://doi.org/10.1038/ncomms11921} {\bibfield
  {journal} {\bibinfo  {journal} {Nature Communications}\ }\textbf {\bibinfo
  {volume} {7}},\ \bibinfo {pages} {11921} (\bibinfo {year}
  {2016})}\BibitemShut {NoStop}%
\bibitem [{\citenamefont {Giorgianni}\ \emph {et~al.}(2019)\citenamefont
  {Giorgianni}, \citenamefont {Cea}, \citenamefont {Vicario}, \citenamefont
  {Hauri}, \citenamefont {Withanage}, \citenamefont {Xi},\ and\ \citenamefont
  {Benfatto}}]{Lara19}%
  \BibitemOpen
  \bibfield  {author} {\bibinfo {author} {\bibfnamefont {F.}~\bibnamefont
  {Giorgianni}}, \bibinfo {author} {\bibfnamefont {T.}~\bibnamefont {Cea}},
  \bibinfo {author} {\bibfnamefont {C.}~\bibnamefont {Vicario}}, \bibinfo
  {author} {\bibfnamefont {C.~P.}\ \bibnamefont {Hauri}}, \bibinfo {author}
  {\bibfnamefont {W.~K.}\ \bibnamefont {Withanage}}, \bibinfo {author}
  {\bibfnamefont {X.}~\bibnamefont {Xi}},\ and\ \bibinfo {author}
  {\bibfnamefont {L.}~\bibnamefont {Benfatto}},\ }\bibfield  {title} {\bibinfo
  {title} {{L}eggett mode controlled by light pulses},\ }\href
  {https://doi.org/10.1038/s41567-018-0385-4} {\bibfield  {journal} {\bibinfo
  {journal} {Nature Physics}\ }\textbf {\bibinfo {volume} {15}},\ \bibinfo
  {pages} {341} (\bibinfo {year} {2019})}\BibitemShut {NoStop}%
\bibitem [{\citenamefont {Jackson}\ \emph {et~al.}(2023)\citenamefont
  {Jackson}, \citenamefont {Dale}, \citenamefont {Maki}, \citenamefont {Xie},
  \citenamefont {Olsen}, \citenamefont {Ahmed-Braun}, \citenamefont {Zhang},\
  and\ \citenamefont {Thywissen}}]{PhysRevX.13.021013}%
  \BibitemOpen
  \bibfield  {author} {\bibinfo {author} {\bibfnamefont {K.~G.}\ \bibnamefont
  {Jackson}}, \bibinfo {author} {\bibfnamefont {C.~J.}\ \bibnamefont {Dale}},
  \bibinfo {author} {\bibfnamefont {J.}~\bibnamefont {Maki}}, \bibinfo {author}
  {\bibfnamefont {K.~G.~S.}\ \bibnamefont {Xie}}, \bibinfo {author}
  {\bibfnamefont {B.~A.}\ \bibnamefont {Olsen}}, \bibinfo {author}
  {\bibfnamefont {D.~J.~M.}\ \bibnamefont {Ahmed-Braun}}, \bibinfo {author}
  {\bibfnamefont {S.}~\bibnamefont {Zhang}},\ and\ \bibinfo {author}
  {\bibfnamefont {J.~H.}\ \bibnamefont {Thywissen}},\ }\bibfield  {title}
  {\bibinfo {title} {Emergent $s$-wave interactions between identical fermions
  in quasi-one-dimensional geometries},\ }\href
  {https://doi.org/10.1103/PhysRevX.13.021013} {\bibfield  {journal} {\bibinfo
  {journal} {Phys. Rev. X}\ }\textbf {\bibinfo {volume} {13}},\ \bibinfo
  {pages} {021013} (\bibinfo {year} {2023})}\BibitemShut {NoStop}%
\bibitem [{\citenamefont {Dale}\ \emph {et~al.}(2024)\citenamefont {Dale},
  \citenamefont {Xie}, \citenamefont {Pond~Grehan}, \citenamefont {Zhang},
  \citenamefont {Maki},\ and\ \citenamefont
  {Thywissen}}]{PhysRevA.110.L051302}%
  \BibitemOpen
  \bibfield  {author} {\bibinfo {author} {\bibfnamefont {C.~J.}\ \bibnamefont
  {Dale}}, \bibinfo {author} {\bibfnamefont {K.~G.~S.}\ \bibnamefont {Xie}},
  \bibinfo {author} {\bibfnamefont {K.}~\bibnamefont {Pond~Grehan}}, \bibinfo
  {author} {\bibfnamefont {S.}~\bibnamefont {Zhang}}, \bibinfo {author}
  {\bibfnamefont {J.}~\bibnamefont {Maki}},\ and\ \bibinfo {author}
  {\bibfnamefont {J.~H.}\ \bibnamefont {Thywissen}},\ }\bibfield  {title}
  {\bibinfo {title} {Emergent $s$-wave interactions in orbitally active
  quasi-two-dimensional fermi gases},\ }\href
  {https://doi.org/10.1103/PhysRevA.110.L051302} {\bibfield  {journal}
  {\bibinfo  {journal} {Phys. Rev. A}\ }\textbf {\bibinfo {volume} {110}},\
  \bibinfo {pages} {L051302} (\bibinfo {year} {2024})}\BibitemShut {NoStop}%
\bibitem [{\citenamefont {Li}\ \emph {et~al.}(2008)\citenamefont {Li},
  \citenamefont {Kelkar}, \citenamefont {Medellin},\ and\ \citenamefont
  {Raizen}}]{Li2008-cy}%
  \BibitemOpen
  \bibfield  {author} {\bibinfo {author} {\bibfnamefont {T.~C.}\ \bibnamefont
  {Li}}, \bibinfo {author} {\bibfnamefont {H.}~\bibnamefont {Kelkar}}, \bibinfo
  {author} {\bibfnamefont {D.}~\bibnamefont {Medellin}},\ and\ \bibinfo
  {author} {\bibfnamefont {M.~G.}\ \bibnamefont {Raizen}},\ }\bibfield  {title}
  {\bibinfo {title} {Real-time control of the periodicity of a standing wave:
  an optical accordion},\ }\href {https://doi.org/10.1364/oe.16.005465}
  {\bibfield  {journal} {\bibinfo  {journal} {Opt. Express}\ }\textbf {\bibinfo
  {volume} {16}},\ \bibinfo {pages} {5465} (\bibinfo {year}
  {2008})}\BibitemShut {NoStop}%
\bibitem [{\citenamefont {Al-Assam}\ \emph {et~al.}(2010)\citenamefont
  {Al-Assam}, \citenamefont {Williams},\ and\ \citenamefont
  {Foot}}]{Al-Assam2010-bo}%
  \BibitemOpen
  \bibfield  {author} {\bibinfo {author} {\bibfnamefont {S.}~\bibnamefont
  {Al-Assam}}, \bibinfo {author} {\bibfnamefont {R.~A.}\ \bibnamefont
  {Williams}},\ and\ \bibinfo {author} {\bibfnamefont {C.~J.}\ \bibnamefont
  {Foot}},\ }\bibfield  {title} {\bibinfo {title} {Ultracold atoms in an
  optical lattice with dynamically variable periodicity},\ }\href
  {https://doi.org/10.1103/PhysRevA.82.021604} {\bibfield  {journal} {\bibinfo
  {journal} {Phys. Rev. A}\ }\textbf {\bibinfo {volume} {82}},\ \bibinfo
  {pages} {021604} (\bibinfo {year} {2010})}\BibitemShut {NoStop}%
\bibitem [{\citenamefont {Ville}\ \emph {et~al.}(2017)\citenamefont {Ville},
  \citenamefont {Bienaim{\'e}}, \citenamefont {Saint-Jalm}, \citenamefont
  {Corman}, \citenamefont {Aidelsburger}, \citenamefont {Chomaz}, \citenamefont
  {Kleinlein}, \citenamefont {Perconte}, \citenamefont {Nascimb{\`e}ne},
  \citenamefont {Dalibard},\ and\ \citenamefont {Beugnon}}]{Ville2017-yy}%
  \BibitemOpen
  \bibfield  {author} {\bibinfo {author} {\bibfnamefont {J.~L.}\ \bibnamefont
  {Ville}}, \bibinfo {author} {\bibfnamefont {T.}~\bibnamefont {Bienaim{\'e}}},
  \bibinfo {author} {\bibfnamefont {R.}~\bibnamefont {Saint-Jalm}}, \bibinfo
  {author} {\bibfnamefont {L.}~\bibnamefont {Corman}}, \bibinfo {author}
  {\bibfnamefont {M.}~\bibnamefont {Aidelsburger}}, \bibinfo {author}
  {\bibfnamefont {L.}~\bibnamefont {Chomaz}}, \bibinfo {author} {\bibfnamefont
  {K.}~\bibnamefont {Kleinlein}}, \bibinfo {author} {\bibfnamefont
  {D.}~\bibnamefont {Perconte}}, \bibinfo {author} {\bibfnamefont
  {S.}~\bibnamefont {Nascimb{\`e}ne}}, \bibinfo {author} {\bibfnamefont
  {J.}~\bibnamefont {Dalibard}},\ and\ \bibinfo {author} {\bibfnamefont
  {J.}~\bibnamefont {Beugnon}},\ }\bibfield  {title} {\bibinfo {title} {Loading
  and compression of a single two-dimensional {B}ose gas in an optical
  accordion},\ }\href {https://doi.org/10.1103/PhysRevA.95.013632} {\bibfield
  {journal} {\bibinfo  {journal} {Phys. Rev. A}\ }\textbf {\bibinfo {volume}
  {95}},\ \bibinfo {pages} {013632} (\bibinfo {year} {2017})}\BibitemShut
  {NoStop}%
\bibitem [{\citenamefont {Fischer}\ and\ \citenamefont
  {Parish}(2013)}]{Fischer2013-is}%
  \BibitemOpen
  \bibfield  {author} {\bibinfo {author} {\bibfnamefont {A.~M.}\ \bibnamefont
  {Fischer}}\ and\ \bibinfo {author} {\bibfnamefont {M.~M.}\ \bibnamefont
  {Parish}},\ }\bibfield  {title} {\bibinfo {title} {{BCS-BEC} crossover in a
  quasi-two-dimensional fermi gas},\ }\href
  {https://doi.org/10.1103/PhysRevA.88.023612} {\bibfield  {journal} {\bibinfo
  {journal} {Phys. Rev. A}\ }\textbf {\bibinfo {volume} {88}},\ \bibinfo
  {pages} {023612} (\bibinfo {year} {2013})}\BibitemShut {NoStop}%
\bibitem [{\citenamefont {Zhou}\ \emph {et~al.}(2023)\citenamefont {Zhou},
  \citenamefont {Shi}, \citenamefont {Liu}, \citenamefont {Hu},\ and\
  \citenamefont {Zhang}}]{Zhou2023-sm}%
  \BibitemOpen
  \bibfield  {author} {\bibinfo {author} {\bibfnamefont {J.}~\bibnamefont
  {Zhou}}, \bibinfo {author} {\bibfnamefont {T.}~\bibnamefont {Shi}}, \bibinfo
  {author} {\bibfnamefont {X.-J.}\ \bibnamefont {Liu}}, \bibinfo {author}
  {\bibfnamefont {H.}~\bibnamefont {Hu}},\ and\ \bibinfo {author}
  {\bibfnamefont {W.}~\bibnamefont {Zhang}},\ }\bibfield  {title} {\bibinfo
  {title} {{BCS–BEC} crossover in a quasi-two-dimensional fermi superfluid},\
  }\href {https://doi.org/10.1088/1367-2630/ace98f} {\bibfield  {journal}
  {\bibinfo  {journal} {New J. Phys.}\ }\textbf {\bibinfo {volume} {25}},\
  \bibinfo {pages} {083001} (\bibinfo {year} {2023})}\BibitemShut {NoStop}%
\bibitem [{\citenamefont {Dyke}\ \emph {et~al.}(2016)\citenamefont {Dyke},
  \citenamefont {Fenech}, \citenamefont {Peppler}, \citenamefont {Lingham},
  \citenamefont {Hoinka}, \citenamefont {Zhang}, \citenamefont {Peng},
  \citenamefont {Mulkerin}, \citenamefont {Hu}, \citenamefont {Liu},\ and\
  \citenamefont {Vale}}]{Dyke2016-mi}%
  \BibitemOpen
  \bibfield  {author} {\bibinfo {author} {\bibfnamefont {P.}~\bibnamefont
  {Dyke}}, \bibinfo {author} {\bibfnamefont {K.}~\bibnamefont {Fenech}},
  \bibinfo {author} {\bibfnamefont {T.}~\bibnamefont {Peppler}}, \bibinfo
  {author} {\bibfnamefont {M.~G.}\ \bibnamefont {Lingham}}, \bibinfo {author}
  {\bibfnamefont {S.}~\bibnamefont {Hoinka}}, \bibinfo {author} {\bibfnamefont
  {W.}~\bibnamefont {Zhang}}, \bibinfo {author} {\bibfnamefont {S.-G.}\
  \bibnamefont {Peng}}, \bibinfo {author} {\bibfnamefont {B.}~\bibnamefont
  {Mulkerin}}, \bibinfo {author} {\bibfnamefont {H.}~\bibnamefont {Hu}},
  \bibinfo {author} {\bibfnamefont {X.-J.}\ \bibnamefont {Liu}},\ and\ \bibinfo
  {author} {\bibfnamefont {C.~J.}\ \bibnamefont {Vale}},\ }\bibfield  {title}
  {\bibinfo {title} {Criteria for two-dimensional kinematics in an interacting
  {F}ermi gas},\ }\href {https://doi.org/10.1103/PhysRevA.93.011603} {\bibfield
   {journal} {\bibinfo  {journal} {Phys. Rev. A}\ }\textbf {\bibinfo {volume}
  {93}},\ \bibinfo {pages} {011603} (\bibinfo {year} {2016})}\BibitemShut
  {NoStop}%
\bibitem [{\citenamefont {Cyrot}(1973)}]{Cyrot1973-oe}%
  \BibitemOpen
  \bibfield  {author} {\bibinfo {author} {\bibfnamefont {M.}~\bibnamefont
  {Cyrot}},\ }\bibfield  {title} {\bibinfo {title} {Ginzburg-landau theory for
  superconductors},\ }\href {https://doi.org/10.1088/0034-4885/36/2/001}
  {\bibfield  {journal} {\bibinfo  {journal} {Reports on Progress in Physics}\
  }\textbf {\bibinfo {volume} {36}},\ \bibinfo {pages} {103} (\bibinfo {year}
  {1973})}\BibitemShut {NoStop}%
\end{thebibliography}%




\clearpage
\pagebreak
\begin{center}
\textbf{\large Supplementary Material: Hybridization of the amplitude mode in a confined fermionic superfluid}
\end{center}

\setcounter{figure}{0}
\renewcommand{\thefigure}{S\arabic{figure}}


\section{\label{sec:prep}Preparation}
\subsection{Box potential}
We prepare a homogeneous Fermi gas of $^6$Li atoms in the lowest two hyperfine states $\ket{1}$ and $\ket{2}$. 
A Feshbach resonance located at $832\,\mathrm{G}$ allows to control the interaction strength. 
The confinement in the $xy$-plane is provided by a radial repulsive optical ring potential with diameter $d=\SI{86}{\micro m}$ \cite{Hueck2018-cs}.
In the vertical direction, the gas is strongly confined to a single layer of a blue-detuned lattice potential with tunable spacing ($\SI{20}{\micro m}$ - $\SI{3}{\micro m}$) \cite{Li2008-cy,Al-Assam2010-bo, Ville2017-yy}.
This results in trapping frequencies of up to $\omega_z=2 \pi\cdot\SI{12}{kHz}$. 

A typical density is $n_{\mathrm{2D}}\approx\SI{1.5}{atoms/\micro m^2}$, corresponding to a Fermi energy $E_\mathrm{F}\approx h\cdot\SI{15}{kHz}$ and Fermi temperature $T_\mathrm{F}\approx\SI{700}{nK}$. 
The lowest temperature in our system is $T^* \approx \SI{20}{nK} \approx 0.014 T_\mathrm{F}$, which is extracted from a wing fit of the non-condensed fraction after T/4 time-of-flight (ToF) expansion in the BEC regime ($B=\SI{750}{G}$). 
This system is the starting point for the experiments presented in the main text, whose parameters are summarized in Tab.\,\ref{tab:parameters}.

\begin{table*}[tb]
    \centering
    \begin{tabular}{|l|c|c|c|c|c|}
         \hline
         Fig.\ & avg.\ & n $[\si{1/\micro m^2}]$ & $\omega_\mathrm{F}/2\pi [\si{kHz}]$ & $\omega_z/2\pi [\si{kHz}]$ & $\omega_\mathrm{F}/\omega_Z$\\ \hline
         2\,A \& 3\,A ($\omega_z$-mod)& 10 & $1.54\pm0.07$ & $16.2\pm0.4$ & $8.57\pm0.5$ & $1.89\pm0.11$\\
         2\,A (onset $2\Delta$) & 20 & $1.19\pm0.05$ & $12.5\pm0.6$ & $6.15\pm0.04$ & $2.03\pm0.02$\\
         3\,B (oscillations) & 5 & $1.46\pm0.14$ & $15.4\pm1.4$ & $8.57\pm0.5$ & $1.80\pm0.11$\\
         4 ($T$-dependence) & 15 & $1.35\pm0.04$ & $14.3\pm0.4$ & $11.46\pm0.33$ & $1.25\pm0.04$\\
         \hline
    \end{tabular}
    \caption{Number of averages, densities, Fermi energies and trapping frequencies for the measurements displayed in the referenced figures.}
    \label{tab:parameters}
\end{table*}

\subsection{2D Parametrization}

In our experiments, the Fermi energy $E_\mathrm{F}$ is comparable to the trap level spacing $\hbar \omega_z$.
This motivates our parametrization of the system in terms of pure 2D quantities using the column density $n_{2D}$, 2D Fermi energy $E_\mathrm{F} = \hbar^2 k_\mathrm{F}^2/2 m$, Fermi momentum $k_\mathrm{F} = \sqrt{4\pi n_{\mathrm{2D}}}$, and the atom mass $m$.

In the presence of strong confinement the quasi-2D scattering amplitude is given by 
\begin{equation}
    f_{\rm{\mathrm{quasi,2D}}}(k,a_{\rm{3D}},l_z) = \frac{4\pi}{\sqrt{2\pi} l_z/a_{\rm{3D}} + w(k^2 l_z^2/2)},
    \label{eq:quasi2DScatteringAmplitude}
\end{equation}
with the 3D scattering length $a_{\mathrm{3D}}$, the harmonic oscillator length $l_z=\sqrt{\hbar/(m\omega_z)}$ and the momentum-dependent contributions to the scattering amplitude $w(x)$ \cite{Petrov2001-kn}. 
In the low-momentum limit $k l_z \rightarrow 0$, the quasi-2D scattering amplitude simplifies to  

\begin{equation}
    f_{\rm{quasi,2D,0}}(k,a_{\rm{3D}},l_z) = \frac{4\pi}{\sqrt{2\pi} l_z/a_{\rm{3D}} - \mathrm{ln}\left(\frac{2\pi}{A} \frac{k^2 l_z^2}{2}\right) + i\pi} 
    \label{eq:quasi2DLowKScatteringAmplitude}
\end{equation}
\\
with $A\approx 0.915$.
Comparing the pure 2D scattering amplitude with the simplified quasi-2D scattering amplitude in Eq.\,\ref{eq:quasi2DLowKScatteringAmplitude} it is possible to obtain a relation between the 2D scattering length and the 3D scattering length,
\begin{equation}
    a_{\mathrm{2D}} = l_z \sqrt{\frac{\pi}{A}} e^{-\sqrt{\frac{\pi}{2}} l_z/a_{\rm{3D}}}
    \label{eq:a2D0}
\end{equation}

The 2D scattering length is used in the main text to define the interaction 
strength $\eta = \ln(k_\mathrm{F} a_\mathrm{2D})$ 
and the mean-field pairing gap
\begin{equation}
    \Delta = \sqrt{2 E_\mathrm{b} E_\mathrm{F}} 
    \label{eq:deltaMF}
\end{equation}
with the binding energy of the 2D bound state $E_\mathrm{b} =  \hbar^2/(m a_{\rm 2D}^2)$.
This parametrization enables us to compare the experimental data with the prediction from an effective field theory, whose free parameters are constrained by existing mean-field and Quantum Monte Carlo calculations. For more details, see Section \ref{sec:effectiveTheory}. 

\begin{figure}[tb!]
    \centering
    \includegraphics[width=\linewidth]{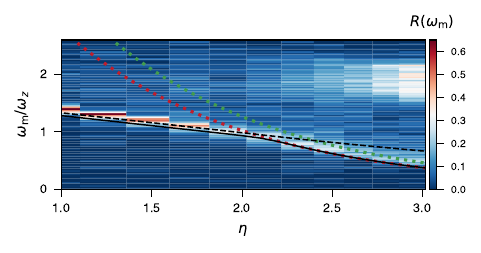}
    \caption{Comparison of the spectrum shown in Fig.\,2 in the BCS regime with the 2D pairing gap (red dashed) and pairing gap taking the vertical confinement into account (Eq.\,\ref{eq:quasi2D}, green dashed). In addition, the resonance positions from the effective field theory are shown (black dashed and black dashed lines, see Fig.\,2).}
    \label{fig:S6}
\end{figure}

We note that beyond mean-field effects as well as the occupation of excited states along the confined direction can lead to modifications of $\Delta$~\cite{Fischer2013-is,Zhou2023-sm}. 
When considering only the lowest two  harmonic states $n = 0, 1$, an analytical expression is given by~\cite{Fischer2013-is} 
\begin{equation}
\frac{\Delta}{E_\mathrm{F}} \simeq \sqrt{\frac{2 E_\mathrm{b}}{E_\mathrm{F}}}\left(1+\frac{E_\mathrm{F}}{8 \omega_z}\right),
\label{eq:quasi2D}
\end{equation}
which we plot in Fig.\,\ref{fig:S6} (green dashed line).

The figure suggest that taking into account the occupation of higher levels should improve the agreement of the effective field theory with the experimental data.
As accurate predictions for speed of sound are not available for our ratio of $E_\mathrm{F}/\omega_z$, we choose to consistently use purely 2D predictions, where QMC calculations are available.

\section{Probing}

\subsection{\label{sec:response} Response measurement}

\begin{figure}[t!]
    \centering
    \includegraphics[width=\linewidth]{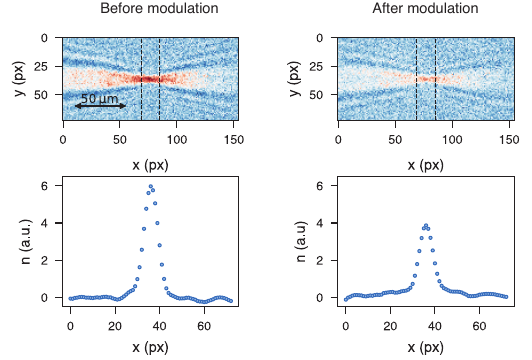}
    \caption{Momentum distribution after trap modulation is acquired by expanding the gas in an in-plane harmonic trap for a quarter trap period. 
    Due to the strong vertical confinement, the gas expands rapidly out of the depth of field of the high-resolution microscope. 
    Therefore, we tilt the imaging axis, to map the z-position to the x-axis and extract the in-focus part (region withing black dashed lines). 
    The in-focus momentum distributions shows a clear condensate peak. 
    Its reduction allows for the extraction of the deposited heat during modulation.
    }
    \label{fig:S1}
\end{figure} 

Trap modulation spectroscopy is performed after creating a strongly confined and homogeneous system. 
At a given interaction strength $\eta$, the trapping frequency is modulated by $\SI{0.5}{\percent}$ at a given frequency $\omega_{\textrm{m}}$ (see Fig.\,1\,c in the main text).
This scheme allows a deposition of energy into the system when being resonant with a mode, resulting in an increase of temperature and subsequent reduction of the condensate fraction after a rethermalization time of $\SI{20}{\ms}$.

In order to extract the response, the reduction in the condensate amplitude $A(\omega_{\textrm{m}})$ from the bimodal momentum distribution is used. 
We define the response as $R(\omega_{\textrm{m}})=A(0)/A(\omega_{\textrm{m}})-1$, which is a proxy for the deposited heat into the system \cite{Biss2022-ra}. 
The amplitude $A(\omega_{\textrm{m}})$ is extracted by mapping to momentum space via a ToF expansion for a quarter period in an in-plane harmonic potential \cite{Hueck2018-cs}.
Due to the strong vertical confinement, the cloud simultaneously expands rapidly along z, such that it becomes larger than the depth of field of the high-resolution microscope.
Therefore, we tilt the imaging beam with respect to the optical axis of the imaging system.
This projects the z-position onto the x-axis, which allows us to extract the in-focus momentum distribution (Fig.\,\ref{fig:S1}) with a horizontal average over the central region within the black-dashed lines.
To obtain the condensate peak height, we perform a moving average of $\SI{8}{px}$ over the momentum distribution and the take the peak density $A_i(\omega_{\textrm{m}})$.
For each modulation frequency $\omega_{\textrm{m}}$, we take multiple images, extract the peak densities $A_i(\omega_{\textrm{m}})$ and average them to $A(\omega_{\textrm{m}})$. 

\begin{figure}[!tb]
\includegraphics[width=\linewidth]{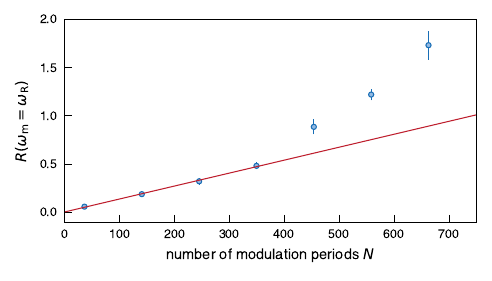}
\caption{The response of modulation spectroscopy for varying number of modulation periods $N$. 
The response increases linearly up to a $N=350$ modulation periods, shown by linear fit taking the first four points into account. 
For these measurements a modulation amplitude $\alpha=1\,\%$ and frequency $\omega_\mathrm{m} = 18 \,\mathrm{kHz}$ is used.} 
\label{SFig:7}
\end{figure}

To ensure linear response, we drive the system on resonance ($\SI{18}{kHz}$ at $\SI{900}{G}$) for different numbers of modulation periods $N$ and extract the response, shown in Fig.\,\ref{SFig:7}. The response behaves linearly up to $N=350$ oscillation periods. We want to stay well below this point and therefore fix the modulation periods to $N=240$ for $\omega_{\textrm{m}} > \SI{2}{\kilo\Hz}$. 
Below $\SI{2}{\kilo\Hz}$, we keep the modulation time at $240/\SI{2}{kHz}=\SI{120}{\milli\second}$, to avoid diverging probe times when going to low frequencies. 
All resonances  we observe via trap modulation are well above $\SI{2}{kHz}$.

\subsection{\label{sec:dynamics} Oscillation Measurement}

The measurements of the time dynamics presented in Fig.\,3b in the main text use a similar protocol presented in section\,\ref{sec:response}.
Here, the system is excited with a short modulation of only two oscillation periods, see inset Fig.\,3\,b, with a large modulation amplitude of $\alpha = \SI{3.5}{\percent}$. 
Afterwards, the system evolves freely for different times $t$, where the occupation of the $n_z=2$ harmonic oscillator state periodically oscillates.
When the vertical confinement is suddenly removed, the excess of transverse kinetic energy due to the partial population of the $n_z=2$ state results in a rapid expansion along $z$ leading to  a larger transverse cloud width and a rapid decrease in the central density of the cloud.
The time evolution of the central density is detected after a short ToF of $\SI{3}{\ms}$.
A similar technique has been used in Ref.\,\cite{Dyke2016-mi} to measure the transverse excitations in quasi-2D Fermi gases.
We fit the observed oscillations with an exponential decay $n_0(t) \propto \mathrm{sin} (\omega_0 t+\phi_0)e^{-t/\tau}$ with decay time $\tau$ and phase $\phi_0$.
From the fitted decay time we obtain the quality factor $Q = \tau \omega_0/2$.

\subsection{\label{sec:temperature} Temperature dependence}

\begin{figure}[t!]
    \centering
    \includegraphics[width=\linewidth]{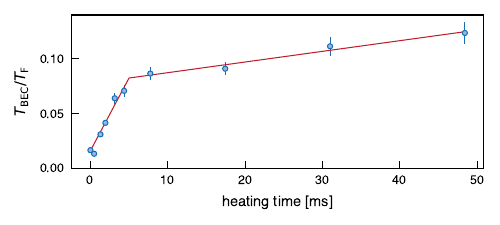}
    \caption{Temperature $T^*$ of the atomic cloud measured in the BEC regime. 
    We employ Bragg spectroscopy to locally heat up the system for different driving time. 
    An empirical bilinear fit (red line) is used to obtain a calibration from heating time to temperature.}
    \label{fig:S3}
\end{figure}

To investigate the temperature effects on the order parameter presented in Fig.\,4 in the main text, we heat up the system by exciting the phonon mode at low momentum via Bragg spectroscopy ($q = 0.48\,k_\mathrm{F}$).
After a variable heating time followed by a fixed time for rethermalisation, we are able to have full control of the temperature of the system. 
The temperature, defined as $T^*$ in the main text is calibrated by performing a wing fit to the momentum distribution acquired after T/4 ToF measurement in the BEC regime ($B=750\,\mathrm{G}$ and $\eta=-1.7$) \cite{Hueck2018-cs}.
The temperature $T^*$ as a function of driving time is shown in Fig.\,\ref{fig:S3}. 
To accurately extract the temperature from the thermal wings, each point is averaged over 50 measurements. 
We use an empirical bilinear fit (red line), to translate the heating times to the temperatures presented in Fig.\,4 in the main text.
For the largest temperature of $T^*/T_\mathrm{F} \sim 0.12$, the thermal excitations are still below the first excited stated due to confinement ($\hbar \omega_z / k_\mathrm{B} T^* > 6$). 

\subsection{Extraction of the pair-breaking continuum}

\begin{figure}[t!]
    \centering
    \includegraphics[width=\linewidth]{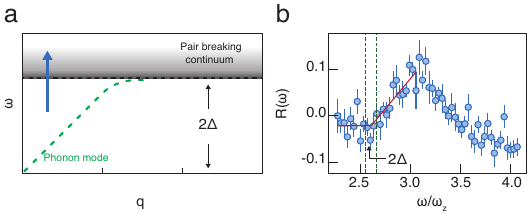}
    \caption{\textbf{a)} 
    Sketch of the excitation spectrum of a fermionic superfluid in the BCS regime. 
    The phononic Bogoliubov mode (dashed green) with the typical linear dispersion relation at low momentum $q$ merges into the pair-breaking continuum (gray area) which is gapped by $2\Delta$.
    The blue arrow indicates the regime where Bragg spectroscopy is performed to extract the pairing gap. 
    \textbf{b)}
    Experimental measurement of the pairing gap using Bragg spectroscopy at momentum $q=0.48\, k_\mathrm{F}$.
    The position of the onset is extracted from a bilinear fit (red line).
    It is shown with its corresponding uncertainty in position (green dashed lines).
    }
    \label{fig:S2}
\end{figure}

To compare the position of the gap with respect to the order parameter oscillations, we perform Bragg spectroscopy with low in-plane momentum $q = 0.48\,k_\mathrm{F}$ (dashed line in Fig.\,\ref{fig:S2}\,a).
For Fig.\,2 in the main text, we determine the size of the pairing gap by finding the onset of the pair-breaking continuum (red points) located at $2\Delta$. 
The spectral response of the onset is shown in Fig.\,\ref{fig:S2}\,b (blue points). 
The gap is extracted from the intersection point of a bilinear fit (red line).

\section{\label{sec:effectiveTheory} Effective field theory}

To gain further insights, we model the system using a low-energy effective field theory \cite{Pekker2015-lk,Skulte2021} describing the dynamics of a complex-valued bosonic scalar field $\Psi$ which represents the condensed pairs. 
To be able to capture both, the BCS- and BEC-dynamics, we include a dynamical term in the form of $ K_2 |\partial_t \Psi(\mathbf{r},t)|^2$ to model particle-hole symmetric dynamics and a term of the form of $i K_1 \Psi^*(\mathbf{r},t) \partial_t \Psi(\mathbf{r},t)$ to include non-particle-hole symmetric dynamics. The parameters $K_{1}$ and $K_{2}$ control the influence of both terms and are determined by the underlying system as discussed later.
The action can be written as
\begin{equation}
\mathcal{S} = \mathcal{S}_\mathrm{dyn}+\mathcal{S}_\mathrm{stat} \equiv \int dt \mathcal{L}.
\end{equation} 
We use the following action and Lagrangian in the form of

\begin{align}
\label{eq:actionDyn}
   &\mathcal{S}_\mathrm{dyn}=\\ &\int \mathrm{d}^3\mathrm{r}\mathrm{d}\mathrm{t}~\bigg( i K_1 \left(\Psi(\mathbf{r},t) \partial_t \Psi^*(\mathbf{r},t)-\mathrm{c.c.} \right) + K_2 |\partial_t \Psi(\mathbf{r},t)|^2 \bigg) \notag 
\end{align}
\begin{align}
\label{eq:actionStat}
    &\mathcal{S}_\mathrm{stat}=\\ &\int \mathrm{d}^3\mathrm{r}\mathrm{d}\mathrm{t}\bigg(  -r_0 |\Psi(\mathbf{r},t)|^2 + \frac{u_0}{2} |\Psi(\mathbf{r},t)|^4 + \frac{\hbar^2}{2 M} |\nabla \Psi(\mathbf{r},t)|^2 \bigg) \notag
    \label{lagrangedensity}
\end{align}
where the parameters $K_{1}$ and $K_2$ interpolate between particle-hole symmetric and non particle-hole symmetric dynamics, $r_0$ is a bias with the character of a chemical potential and $u_0$ is the density-density interaction strength. The precise values of these parameters have to be determined for different values of $\eta$ throughout the crossover. We note that the static part only contains standard terms included in Ginzburg-Landau theory~\cite{Cyrot1973-oe}. \\
Due to the second time derivative, the total particle number is not conserved as we can see from the Noether charge, which is given by
\begin{equation}
J = i \int \mathrm{d}^3r \left[ \Pi (r,t) \Psi  (r,t)-\Pi (r,t)^* \Psi (r,t)^* \right]
\end{equation}
with the canonical conjugate defined as 
\begin{equation}
\Pi(r,t) = \frac{\partial \mathcal{L}}{\partial (\partial_t  \Psi (r,t))} = K_2 \partial_t \Psi^*(r,t)- i K_1 \Psi^*(r,t).
\end{equation}

\subsection{Determination of the free parameters}
In order to extract predictions from the effective field theory, the values of our free parameters $K_{1/2}$, $u_0$, $M$ and $r_0$ have to be determined throughout the entire BEC-BCS crossover. 
We use the following constraints: As the complex scalar field describes the bosonic Cooper pair wavefunction, we set $M=2 m$. We will further consider that during the crossover the energy scale has to be fixed. This enforces a constraint on the sum of $K_1$ and $K_2$. 
To recover the GPE limit for $K_2=0$ and the BCS limit for $K_1=0$ we choose
\begin{equation}
K_1 = \hbar - 2 K_2  E_\mathrm{F} / \hbar.
\label{K1K2relation}
\end{equation}
Next, we identify the amplitude and Goldstone mode in our effective model in the absence of confinement. For this we invoke the Euler-Lagrange equations to obtain
\begin{multline}
\partial_t \left(-K_2 \partial_t+i K_1 \right) \Psi (r,t) =\\ (-r_0-\frac{\hbar^2}{2M}\nabla^2+u_0 |\Psi(r,t)|^2) \Psi (r,t).
\end{multline}
We obtain the steady state solution by setting all time derivatives to zero and find $|\Psi_0|^2=\frac{r_0}{u_0}$. Next, we expand our complex field around the fixed point to lowest order \begin{equation}
\label{SEq:1}
\Psi = \Psi_0+\delta_a + i \Psi_0 \phi \equiv \Psi_0+\delta_a + i \delta_\phi,
\end{equation}
where we used the underlying $U(1)$ symmetry to choose the phase of the fixed point to be zero. We expand the complex field in Eq.~\ref{SEq:1}, keeping only the lowest order. By taking the Fourier transform and separating the real and imaginary part we obtain
\begin{align}
 K_2 \omega^2  \delta_a- i \omega K_1 \delta_\phi&=2r_0\delta_a+\frac{\hbar^2}{2M}q^2 \delta_a \\
 K_2 \omega^2  \delta_\phi+ i \omega K_1 \delta_a&=\frac{\hbar^2}{2M}q^2 \delta_\phi,
\end{align} where $\omega$ is the frequency and $q$ the momentum.
Solving for $\omega$, we obtain
\begin{widetext}
\begin{align}
\omega_{\phi}=\sqrt{\frac{K_1^2+2K_2\left( \frac{\hbar^2}{2M}q^2+r_0 \right)-\sqrt{K_1^4+4K_2^2 r_0^2 + 4K_1^2K_2\left( \frac{\hbar^2}{2M}q^2+r_0 \right)}}{2K_2^2}} \label{eq:omegaPhi}\\
\omega_{a}=\sqrt{\frac{K_1^2+2K_2\left( \frac{\hbar^2}{2M}q^2+r_0 \right)+\sqrt{K_1^4+4K_2^2 r_0^2 + 4K_1^2K_2\left( \frac{\hbar^2}{2M}q^2+r_0 \right)}}{2K_2^2}},
\end{align}
\end{widetext}
where $\omega_a$ denotes the amplitude mode and $\omega_\phi$ the Goldstone mode.
For the special case $K_1=0$, corresponding to  particle-hole symmetry, we recover the well-known result, that the modes decouple~\cite{Pekker2015-lk} and are given by
\begin{align}
\omega_{\phi,K_1=0}=\sqrt{\frac{\hbar^2}{2MK_2}}|q| \\
\omega_{a,K_1=0}=\sqrt{\frac{2r_0+\frac{\hbar^2}{2M}q^2}{K_2}}.
\end{align}
Taking the limit of $K_2 \rightarrow 0$ in Eq.~\ref{eq:omegaPhi} recovers the Boguliubov dispersion relation with $K_1=\hbar$ and $r_0=\mu$
\begin{equation}
\hbar \omega=\sqrt{\left(\frac{\hbar^2 q^2}{2M}\right)^2+2 \mu \frac{\hbar^2 q^2}{2M}}.
\end{equation}

\begin{figure}[!tb]
\hspace*{-0.7cm} 
\includegraphics[width=\linewidth]{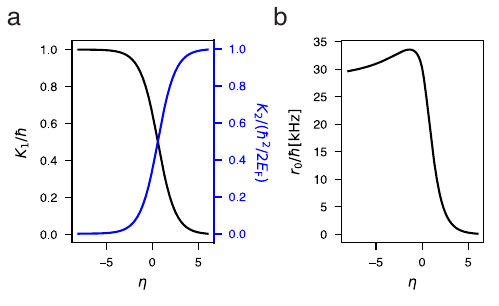}
\caption{The effective parameters in the crossover. \textbf{a)} $K_1$ in units of $\hbar$ and $K_2$ in units of $\hbar^2/(2E_\mathrm{F})$ and \textbf{b)} $r_0$ in units of $\hbar \, \mathrm{kHz}$. Not shown is $u_0$ that can be simply computed by $r_0/n_0$ and follows a similar trend as $r_0$. As $\eta$ flips sign a rapid change in parameters can be observed. } 
\label{SFig:4}
\end{figure}

To determine the remaining free parameters, we use the condensate density $n_0$ in the 2D BEC-BCS crossover, which we take from~\cite{Salasnich2007}. 
For the amplitude mode energy $2\Delta$, we take the 2D mean-field result (Eq.\,\ref{eq:deltaMF}).
Additionally, we consider the speed of sound $v_s$ as reported from previous work using 2D Quantum Monte Carlo ~\cite{Shi2015}.
In our model these quantities relate to each other as
\begin{align}
n_0 &= r_0/u_0 \\
v_s &=\frac{\partial \omega_a}{\partial q} \left. \right|_{q=0}= \hbar^2 \sqrt{\frac{r_0}{M}} \sqrt{\frac{1}{\left( \hbar^2-2 E_\mathrm{F} K_2 \right)^2+2 \hbar^2K_2 r_0}} \\
2 \Delta &= \hbar \omega_a\left.\right|_{q=0}= \sqrt{\frac{2(\hbar-\frac{2 E_\mathrm{F} K_2}{\hbar})^2+4 K_2 r_0}{2K_2^2}},
\end{align}
thereby fixing the remaining free parameters in our effective low-energy model, $r_0$, $u_0$, $K_1$ and $K_2$ (using Eq.~\ref{K1K2relation}). \\ Fig.\,\ref{SFig:4} depicts these parameters between $\eta = -8$ and $\eta=6$ during the crossover. We find that for $\eta < -4$ we recover the expected broken particle-hole symmetry of a BEC, hence $K_2 \approx 0 $ and $K_1 \approx \hbar $. On the BCS side we find that $K_2 \approx \hbar^2/ 2 E_\mathrm{F}$ and $K_1 \approx 0$.

\subsection{Effective two-mode model}
After fixing the parameters by considering a 2-dimensional condensate, we introduce a strong spatial confinement along the $z$ direction analogous to the experiment and expand the wavefunction to find a simplified two-mode model that we use for our numerical results in the main text. We also carry out numerics including the full $z$-dynamics and quantitatively find the same results. For simplicity, and as it allows a direct interpretation of the emerging mode, we have chosen to present the two-mode model here and in the main text.

We start by rewriting the complex scalar field as
\begin{equation}
\Psi(x,y,z) = \Phi(x,y)\Phi(z).
\end{equation}
Since the experiment uses a box potential, we assume that there are no spatial modulations along the $x-y$ direction and expand the field as uniform in this plane. Along the $z$ direction, we have a strong harmonic confinement, meaning the atoms only occupy the harmonic oscillator ground state. Assuming that our effective $u_0$ is small and only acts as a perturbation to the harmonic oscillator ground state and assuming we only access the lowest eigenstates, we expand the complex scalar field as
\begin{equation}
\Psi(x,y,z)  = \frac{1}{L_x L_y} \left[  f_0(z) \alpha_0+f_1(z) \alpha_1+f_2(z) \alpha_2 \right]
\end{equation}
with the harmonic oscillator (HO) eigenfunctions
\begin{equation}
f_n(z) = \frac{1}{\sqrt{2^n n!}}\left( \frac{1}{l_{z,M}\sqrt{\pi}} \right) ^{\frac{1}{2}} \hspace{-1.5ex}e^{-\frac{M\omega_z}{2\hbar}z^2} H\left(n,\frac{z}{l_{z,M}}\right)
\end{equation}
with the harmonic oscillator length $l_{z,M}=\sqrt{\hbar/(M \omega_z)}$ and n-th Hermite polynomial $H(n,z)$. Due to the driving in the experiment, the system can couple to higher states. However, due to parity symmetry, our drive does not couple to the first, but to the second excited state.
By construction, integration over the $x-y$ direction leads to unity. 
We checked numerically that the first level of the HO is always empty during the dynamics and we neglect this contribution in the following, reducing our system to the ground and second excited state. 
Next, we use this expansion to compute the integrals in Eq.~\ref{eq:actionStat} and Eq.~\ref{eq:actionDyn}, from which we obtain
\begin{widetext}
\begin{align}
&\int d^3 r \Psi^* (x,y,z) \biggl[-r_0+\frac{u_0}{2}\Psi^* (x,y,z)\Psi(x,y,z) \biggr] \Psi(x,y,z) = -r_0 (\alpha^*_0 \alpha_0+\alpha^*_2 \alpha_2) \\ \notag & +\frac{u_0}{2} \frac{1}{16}\sqrt{\frac{M \omega_z}{\pi \hbar}} \left[12 \sqrt{2}\alpha^*_2\alpha^*_0\alpha_2\alpha_0+8\sqrt{2}\alpha^*_0\alpha^*_0\alpha_0\alpha_0+\frac{41}{8}\sqrt{2}\alpha^*_2\alpha^*_2\alpha_2\alpha_2\right]  \\ &+\frac{u_0}{2} \frac{1}{16}\sqrt{\frac{M \omega_z}{\pi \hbar}}\left[\left( -8\alpha^*_0\alpha^*_0 \alpha_2\alpha_0+\alpha^*_2\alpha^*_2\alpha_2\alpha_0+3\sqrt{2}\alpha^*_2\alpha^*_2\alpha_0\alpha_0 \right)+\mathrm{h.c.} \right].  \notag
\end{align}
From the driven HO contribution, we further assume $\omega_z(t)^2=\omega_z^2+a(t)$ and $a(t)=b \omega_z^2 \sin(\omega_\mathrm{m}t)$. We compute
\begin{align}
&\int d^3 r \Psi^*(x,y,z) \biggl[ -\frac{\hbar^2}{2M}\partial^2_z+\frac{1}{2}M(\omega_z^2+a(t)) z^2\biggr] \Psi(x,y,z) \\ \notag & =\frac{\hbar \omega_z}{2} \left(\alpha^*_0\alpha_0+5\alpha^*_2\alpha_2+\frac{b(t)}{2}\left[\alpha^*_0\alpha_0+5\alpha^*_2\alpha_2+\frac{\sqrt{2}}{2}\left(\alpha^*_2\alpha_0+\alpha^*_0\alpha_2 \right)  \right] \right),
\end{align}
with the short hand notation $a(t)=b \omega_z^2 \sin(\omega_\mathrm{m}t) = b(t) \omega_z^2$.
Finally, for the dynamical part of the Lagrangian we obtain
\begin{align}
\int d^3 r \mathcal{L}_\mathrm{dyn} = -K_2 \left( \partial_t \alpha_0\partial_t \alpha^*_0+\partial_t\alpha_2\partial_t \alpha^*_2 \right)+i K_1 \left\{ \left( \alpha^*_0\partial_t \alpha_0-\alpha_0\partial_t \alpha^*_0\right)+\left( \alpha^*_2\partial_t \alpha_2-\alpha_2\partial_t \alpha^*_2\right) \right\}.
\end{align}
Invoking the Euler-Lagrange equations then leads to the equations of motion:
\begin{align}
\label{eq:2modeA}
&\partial_t \left(-K_2 \partial_t+i K_1 \right) \alpha_0 = \left(-\tilde r_0+\frac{\hbar \omega_z}{2}\left[1+b(t)\right] \right) \alpha_0 +b(t)\frac{\sqrt{2}\hbar \omega_z}{4} \alpha_2 \\ 
&+\frac{u_0}{2} \frac{1}{16}\sqrt{\frac{M \omega_z}{\pi \hbar}} \left[ 12\sqrt{2} \alpha^*_2 \alpha_2 \alpha_0 + 16 \sqrt{2}\alpha^*_0 \alpha_0 \alpha_0-16 \alpha^*_0 \alpha_2 \alpha_0-8 \alpha^*_2 \alpha_0 \alpha_0+\alpha^*_2 \alpha_2 \alpha_2+6\sqrt{2}\alpha^*_0 \alpha_2 \alpha_2  \right] \notag \\ 
 \label{eq:2modeB}
& \partial_t \left(-K_2 \partial_t+i K_1 \right) \alpha_2 =\left(-\tilde r_0+5\frac{\hbar \omega_z}{2}\left[1+b(t)\right] \right) \alpha_2 +b(t)\frac{\sqrt{2}\hbar \omega_z}{4} \alpha_0 \\
&+\frac{u_0}{2} \frac{1}{16}\sqrt{\frac{M \omega_z}{\pi \hbar}} \left[ 12\sqrt{2} \alpha^*_0 \alpha_2 \alpha_0 + \frac{41}{4} \sqrt{2}\alpha^*_2 \alpha_2 \alpha_2-8 \alpha^*_0 \alpha_0 \alpha_0+2 \alpha^*_2 \alpha_2 \alpha_0+\alpha^*_0 \alpha_2 \alpha_2+6\sqrt{2}\alpha^*_2 \alpha_0 \alpha_0  \right]. \notag
\end{align}
\end{widetext}

\subsection{Numerical results}

We study the linear response of the system to the trap modulation $b(t)$. 
To obtain the black solid line of Fig.\,2\,a in the main text, we proceed as follows.
We choose a driving amplitude $b=0.01$, which is small enough to ensure that we are probing the linear feedback of the system. We further add a small dissipation term in the form of $- \hbar \gamma \partial_t \Psi(z,t)$ in the left-hand side of Eq.\ref{eq:2modeA}-\ref{eq:2modeB} to allow the system to reach a steady state. In the simulation we take $\gamma=10^{-2}$. We numerically solve the order parameter dynamics $\Psi(t)$ following Eq.\,\ref{eq:2modeA}-\ref{eq:2modeB} and extract the Fourier component of $\Delta n(t)=n(t)-n(0)$ at the driving frequency $\omega_\textrm{m}$
\begin{equation}
\Delta n(\omega_\textrm{m})= \int_{t_0}^{t_0+\Delta t} \Delta n(t)  \mathrm{exp} \left( i \omega_\textrm{m} t \right) dt
\end{equation}
where $n(z,t)=\sum_{i=0,2} |\alpha_i|^2$ is the field density.
After the system reaches a steady state at time $t_0$, we consider a time window $\Delta t=50T_\textrm{z}$ ($T_z=2\pi/\omega_z$) which is sufficient to obtain a good frequency resolution.

Finally we compute the response function $\Delta n(\omega_\textrm{m})=\int_{}^{} |\Delta n(z,\omega_\textrm{m})|dz$
for different modulation frequencies $\omega_\textrm{m}$ and extract the frequency $\omega_c$ at which the response takes the maximum value. This frequency $\omega_c$ give us the corresponding collective mode. Doing this for different interaction parameters $\eta=\ln(k_{\mathrm{F}}a_{\mathrm{2D}})$, we obtain the collective mode dependence on $\eta$ which is depicted by a black solid line in Fig.\,2\,a of the main text.

To quantify the amplitude contribution to the collective mode, we define 
\begin{equation}
\kappa_\mathrm{A}= \frac{|\int_{}^{} \Delta n(z,\omega_c)dz|}{\int_{}^{} |\Delta n(z,\omega_c)|dz}, 
\end{equation}
representing the fraction of the total response, which does not conserve the integrated field density.
Following this definition, far in the BEC regime the integral in the numerator vanishes, so $\kappa_\mathrm{A}=0$ and the collective mode has a pure breathing contribution. 
This is because in the BEC regime, the order parameter response is simply redistributed along the $z$ direction in a way that the field density remains constant, resulting in a vanishing numerator after integrating over space.
Conversely, in the BCS regime, the order parameter dynamics is dominated by amplitude oscillations, i.e.\ the integrated field density changes, and the numerator does not vanish.
In this case, the collective mode shows an amplitude contribution ($\kappa_\mathrm{A}>0$).
We compute the amplitude contribution $\kappa_\mathrm{A}$ numerically as a function of $\eta$ which is plotted in Fig.\,2\,b of the main text.

\subsection{Analytical results}
Here we derive the analytical solution for the transition frequency $\Delta\omega=\omega_2-\omega_0$ between the ground and second excited state shown as a dashed black line in Fig.\,2\,a of the main text.

For this we compute the collective modes associated to the bosonic system described by Eqs.\,\ref{eq:2modeA},\ref{eq:2modeB} by keeping only the leading order interacting terms, i.e.\ considering $\alpha_2\sim 0$ and $\alpha_0^*\alpha_0=(r_0-\omega_z K_1/2)/u_0$ in the terms proportional to $u_0$.
By collecting terms proportional to $\alpha_0$ in Eq.\,\ref{eq:2modeA} and proportional to $\alpha_2$ in Eq.\,\ref{eq:2modeB} the equations of motion simplify to:

\begin{align}
&\partial_t \left(-K_2 \partial_t+i K_1 \right) \alpha_0 = \left(-\tilde r_0+\frac{\hbar \omega_z}{2} + \delta_0 \right) \alpha_0,  \\ 
& \partial_t \left(-K_2 \partial_t+i K_1 \right) \alpha_2 = \left(-\tilde r_0+5\frac{\hbar \omega_z}{2} + \delta_2 \right) \alpha_2,
\end{align}
\label{eq:2modesimplified}
with the last terms on the right hand side of both equations representing a correction to the energy (mean-field shift) due to interactions. They are given by:

\begin{align}
&\delta_0=\frac{\sqrt{2}}{2}\sqrt{\frac{M \omega_z}{\pi \hbar}} (r_0-\omega_z K_1/2)\\
&\delta_2=\frac{12\sqrt{2}}{32}\sqrt{\frac{M \omega_z}{\pi \hbar}} (r_0-\omega_z K_1/2).
\end{align}

The collective modes associated to the resulting equations of motion (34) and (35) can be obtained by introducing an ansatz of the type $\alpha_n \sim \exp(i\omega_n t)$. This yields
\begin{equation}
\omega_n^{\pm}=\frac{-K_1}{2 K_2}\pm \sqrt{\frac{K_1^2}{4K_2^2}+\frac{E_n}{K_2}}
\label{eq:omegaNAnalytical}
\end{equation}
with $E_n=\hbar\omega_z (n+\frac{1}{2})-\tilde r_0+\delta_n$ and $n=0,2$.
Finally, we define $\Delta\omega=\omega_2^+-\omega_0^+=-(\omega_2^- -\omega_0^-)$.

\newpage



\end{document}